\documentclass[12pt]{article}
\usepackage{mydef}
\usepackage{authblk}
\usepackage{ctable}
\newcommand{\dd}{\partial}

\usepackage[hide]{ed}

\begin{document}

\title{To sigmoid-based functional description of the volatility smile.}
\author{Andrey Itkin}

\affil{Polytechnic School of Engineering, New York University, \\
6 Metro Tech Center, RH 517E, Brooklyn NY 11201, USA}

\date{\today}

\maketitle
\begin{abstract}
We propose a new static parameterization of the implied volatility surface which is constructed by using polynomials of sigmoid functions combined with some other terms. This parameterization is flexible enough to fit market implied volatilities which demonstrate smile or skew. An arbitrage-free calibration algorithm is considered that constructs the implied volatility surface
as a grid in the strike-expiration space and guarantees a lack of arbitrage at every node of this grid. We also demonstrate how to construct an arbitrage-free interpolation and extrapolation in time, as well as build a local volatility and implied pdf surfaces. Asymptotic behavior of this parameterization is discussed, as well as results on stability of the calibrated parameters are presented. Numerical examples show robustness of the proposed approach in building all these surfaces as well as demonstrate a better quality of the fit as compared with some known models.


\end{abstract}

\section{Overview}

During last 15 years various parameterizations of the implied volatility (IV) surface were proposed in the literature to address few goals: a) given a set of market quotes for some options build an arbitrage-free local volatility (Dupire's) surface to further exploit it for calibration of a local stochastic volatility model; b) obtain volatilities for pricing OTC options and other derivatives with strikes and maturities other than that offered by the option exchanges; c) assess an adequacy of an option pricing model based on the shape of the IV surface. Also option traders and marker makers often use the current snapshot of the implied volatility over all strikes and maturities as a basis to produce a short-term volatility forecast over future periods of time using some assumptions about the future dynamic of the IV\footnote{As was mentioned by one of referees, a single point on the implied volatility surface could potentially be such a forecast. Also market models of implied volatility, e.g., \cite{Cont2002} tell us that implied volatilities also forecast their covariance with spot and their own volatilities as well. However, the usefulness
of even a single implied volatility as a forecast is hampered by the difference between risk-neutral and real world probability measures. It is well known that for S\&P500, the at-the-money forward implied volatility is on average above the subsequent realized volatility, suggesting that this
distinction is important and empirically verifiable. However, in a short run, say up 10 minutes, in a quiet market such a forecast could be potentially helpful.}

There exist two major approaches to construct an arbitrage-free IV surface. The first one uses some stochastic model for the underlying spot or forward price which is calibrated to the market data. For instance, in the equity world one can use the popular Heston (\cite{Heston:1993a}) or SABR (\cite{hagan2002}) model, calibrate it to the market data and then use this model to find the IVs for the missing strikes and expirations where the market quotes are not available. By construction the IVs produced by the arbitrage-free model are also arbitrage-free. However, the main problem with this approach is that it is difficult to come up with a model which is rich enough to fit well the observed market data.

Another interesting rectification of this approach was proposed in \cite{LiptonSepp2011iv} who calibrate a model with tiled local volatility to sparse market data by using the direct and inverse Laplace transforms. The main idea is to have a parametric form for the local volatility with as many parameters as there are market quotes. This allows finding an exact solution to the calibration problem at each forward time step, rather than to solve it in the least-squares sense. So the advantage of the method is that it is pretty fast. On the other hand, as the local volatility is built using an arbitrage-free model, the corresponding IV surface reproduced from the local volatility surface using the Dupire's formula, is arbitrage-free as well.

The other approach does not consider any model of the underlying, but instead uses some parametric fit of the implied volatility surface. Parametric models of the IV came to a regular consideration at the end of 1990s. Several parametric models for the IV surface were suggested by \cite{Dumas98} \footnote{He actually suggested a model for the local volatility, which, however, could be re-mapped to the implied volatility.}, and adapted and tested for FTSE options by \cite{Alentorn04}. In the Dumas parametric model the IV  surface is modeled as a quadratic function of the so-called normalized strike (rather than the strike price). Later this approach was further extended by \cite{Tompkins2001, Kotze2013, Carr2013}. Th normalized strike is defined
\begin{equation} \label{z}
z = \dfrac{\log(K/F)}{\sigma_* \sqrt{T}},
\end{equation}
\noindent where $K$ is the option strike, $F$ is the forward price, $T$ is the time to expiration, and $\sigma_*$ is the normalization constant which usually is set either to 1, or to the ATM implied volatility. The normalized strike is a unit-less quantity. Some people also call it moneyness or log-moneyness, however we reserve this word for a standard definition of the forward moneyness as $M=K/F$. By definition normalized strike vanishes at the forward money (ATM). For a call option, positive normalized strike corresponds to the In-The-Money option and negative normalized strike - to the Out-of-The-Money option. Usually the normalized strike is used under an assumption of "sticky moneyness" which means that the IV doesn't change when $z$ stays constant (it is also known as "sticky delta"), which allows elimination of refitting the volatility smile within some postulated period of time even when the underlying price changes. This is different from another popular assumption which is called a "sticky strike" rule (\cite{Derman1994, Derman1999, Sinclair2013}).

Despite within the second approach (a static parameterization) the quality of the fit is often better than in the first one (a dynamic model of the underlying or the implied volatility itself),
static parameterization tells us something just about the current market snapshot of the option prices/IVs, and nothing about the temporal dynamics of the IV. For instance, the above mentioned "sticky" assumptions about the future dynamics of the IVs are irrelevant to the parameterization itself. Clearly one can construct such a parameterization using the normalized strike as a convenient underlying variable if he/she relies on a "sticky log-moneyness" dynamics assumption to be true.
This, however, doesn't mean that another parameterization which uses the normalized strike as an underlying variable and relies on a sticky-moneyness assumption might not be used to fit the same set of the market IVs. That is because this type of parameterizations is static by nature. In other words, it is impossible to forecast the future IVs {\it per se} using this static fit. Rather, when using this approach by term "forecasting" the practitioners usually mean that when the IV of some option with time horizon (maturity) $T$ is known, it provides some average value of volatility from today to $T$. This, however, is not a property of the parameterization, but rather the property of the current option market to provide some "on average" information about the future behavior of the stock market.

An extended work on modeling the IV surface using the static approach has been done by Gatheral in many papers, starting perhaps with \cite{Gatheral2005}). He used a different parameterization of the smile, known as stochastic-volatility-inspired (SVI) model, which is driven by a forward log-moneyness $\chi = \log(K/F)$. Gatheral and co-workers also proposed some empirical dependencies of how parameters of the fit evolve with time, \cite{Gatheral2006}. Later in \cite{GatheralJacquier2011} it was shown that the SVI parameterization and the large-time asymptotic of the Heston implied volatility agree algebraically, which provides an additional theoretical justification for the above parameterization.

Some other static parameterizations were also proposed in the literature, e.g. \cite{Fengler2005,  ZhaoHodges2013, Andreou2014, Sehgal2008, Daglish2007, Carr2013, Romo2011, Rosenberg2000}, and also references therein. In the next section we discuss the main highly desirable features that any such a parameterization should provide the user with. It could be observed that in contrast to the old approaches, recent models, e.g. the extended SVI model, and models in \cite{Kotze2013,ZhaoHodges2013} do make account for these features, and thus could be useful in practice.

As far as a demand for the dynamic models of the IV is concerned, \cite{Cont2002} considered the prices of the index options at a given date (they are usually represented via the corresponding IV surface) that clearly demonstrated skew/smile features and also a term structure, the behavior that several IV models have attempted to reproduce. They underlined that the IV surface also changes dynamically over time in a way that is not taken into account by the existing modeling approaches, giving rise to a Vega risk in option portfolios. Using time series of option prices on the S\&P500 and FTSE indices, they studied the deformation of this surface and showed that it may be represented as a randomly fluctuating surface driven by a small number of orthogonal random factors. Then Cont and Fonseca identified and interpreted the shape of each of these factors, studied their dynamics and their correlation with the underlying index. A simple factor model compatible with the empirical observations was proposed. The authors illustrated how this approach simulates and improves the well-known "sticky moneyness" rule used by option traders for updating the IVs. Their approach gave justification for using Vega when measuring the volatility risk, and provided decomposition of the volatility risk as a sum of contributions from empirically identifiable factors.

It is worth mentioning that the popular assumptions of "sticky strike" or "sticky moneyness" are just an empirical rule-of-thumb. For instance, \cite{Bouchaud2008} analyzed these assumptions by considering in detail the skew of some stock option smiles, which is induced by the so-called leverage effect on the underlying, i.e., the correlation between past returns and future square returns. This naturally explains the anomalous dependence of the skew as a function of the option maturity. The market cap dependence of the leverage effect is analyzed using a one-factor model. The authors show how this leverage correlation gives rise to a non-trivial smile dynamics, which turns out to be intermediate between the "sticky strike" and the "sticky delta" rules. Finally, they compare their result with stock options data, and find that the option markets overestimate the leverage effect by a large factor, in particular, for the long-dated options. This subject requires some further investigation.

Another interesting idea was proposed in \cite{CarrWu2010}. This paper considers the future dynamics of the Black-Scholes implied volatility surface, and derives no-arbitrage constraints on the current shape of the volatility surface. Under the specified proportional volatility dynamics, the shape of the surface can be cast as solutions to a simple quadratic equation. Furthermore, corresponding to the option implied volatility for each contract, the paper defines a new, option-specific expected volatility measure that can be estimated from the historical sample price path of the underlying security. The measure is defined as the volatility input that generates zero expected delta-hedged gains from holding this option and can thus differ across different option strikes and expiries. Applying the new theoretical framework to the S\&P500 index options market, the authors extract volatility risk and volatility risk premium from the two volatility surfaces, and find that the extracted volatility risk premium significantly predicts future stock returns. Thus, knowledge of the future dynamics also eliminates the necessity in any artificial assumptions like "stickiness", etc. See, a recent paper of \cite{Sepp2014} and also \cite{Roux2007, Romo2014}.

So far, most of the IV researchers have been focused on Equity and FX derivatives. However, \cite{Borovkova2008} applied this idea to the option price data from oil markets. They combined the simplicity of the Gatheral parametric method with the flexibility of a non-parametric approach. The authors claim that the method can successfully deal with a limited amount of the option price data. Performance of the method was investigated by applying it to prices of the exchange-traded crude oil and gasoline options, and the results were compared with those obtained by a purely parametric approach. Furthermore, investigation of the relationship between volatilities implied from the European and Asian options showed that the Asian options in oil markets are significantly more expensive than the theoretical arguments imply.

To summarize, various static parameterizations were in use by traders since 1990 when the skew
became pronounced in the market. However, as practitioners observed in their day-to-day trading, even the best models such as SVI and recent versions of the quadratic fit sometimes fail to fit well the market data. The author's own experience also justifies a failure to fit these models to the data sets, obtained from some data providers. Also, according to \cite{BiscampPC2008} the SVI model was thoroughly tested by practitioners in recent years and did not prove to work well for all products (like the index options, dispersion, equity options etc.). Therefore, some trading firms run their own proprietary models that exploit an idea of building a piecewise polynomial smile in the $z$ space. This approach also has some problems, namely:
\begin{itemize}
\item Determining a boundary point between two pieces of the smile, where in addition the smile is $C_2$ continuous. Usually it requires solving some non-linear equation, which is expensive. The necessity of solving the nonlinear equation slows down the volatility smile fit, and especially computation of derivatives of the smile with respect to the model parameters which usually are computed by the bump-and-grind method.

\item This approach still does not resolve the problem of fitting maturities close to expiration.

\item This functional form does not fit the market data well for both skew and smile.

\item The asymptotic behavior of the smile at wings in $z$ does not agree with the result of \cite{Lee2004} that the variance should be asymptotically linear in $z$.
\end{itemize}

All the above suggests that a new model suitable to better fit the static market volatility data could be helpful.

With allowance for the above our main goal in this paper is to propose another parametric fit which amounts to resolving the discussed issues with the existing approaches. We also show how to construct a arbitrage-free IV surface by using an arbitrage-free interpolation and/or extrapolation if necessary.

We emphasize that according to \cite{PCarr2014} any such a formula must provide the following three properties:
\begin{enumerate}
\item It analytically describes implied volatilities instead of option prices.
\item It exactly fits any set of arbitrage-free mid-market implied volatilities.
\item It does not produce arbitrage.
\end{enumerate}
Similar thoughts could be found in \cite{Rebonato2004, Castagna2010}.

While the first one is obvious, it is usually hard to guarantee the last two properties. In the approach of this paper first, we don't guarantee an exact fit to the given mid-market quotes since we use a least-square optimization. However, we do guarantee, that the regressed implied volatility is in between of the given bid and ask, and is close (in some norm) to the mid price. Second, by construction we guarantee no-arbitrage in time. We also guarantee no-arbitrage on a given grid of strikes. This grid could be non-uniform, and it consists of the nodes with different strikes and/or maturities. However, we don't suggest the arbitrage-free interpolation/extrapolation in the strike space. Therefore, definition of the reasonable grid of strikes is left up to the user of this approach.

As an example, suppose we use the local stochastic volatility model to price and hedge a set of exotic and vanilla options simultaneously. To do that we need a local volatility (LV) surface calibrated to the market data. The appropriate LV grid could, e.g., coincide with the finite-difference grid in the spot space. To get the LV surface we may first build the IV surface and calibrate it to the vanilla quotes, and the use the Dupire's formula to re-map the IV into the LV. When using such an approach we are not interesting in the values of the implied volatilities in between the grid nodes, and therefore, the proposed method could be applied. We also guarantee a correct asymptotic behavior of the smile at both large positive and negative normalized strikes.

The above mentioned means that our model of implied volatility is a {\it discrete} model defined at a given given set of "states" (strikes), similar to, say, a discrete Markov chain model. And we are not aware of any continuous limit of this model at the moment. Compare this with the SVI model where a nice result is available that the model structurally coincides with the hight $T$ asymptotic of the Heston model.

The rest of the paper is organized as follows. In section~\ref{model} we give a general outline of the model. Next section provides an asymptotic analysis of the model behavior at extreme strikes and expirations as well as ATM and some critical strikes. Based on this analysis we also are able to give a financial interpretation of some model parameters. In section~\ref{na} we explain how to construct an arbitrage-free IV surface and describe in detail our approach to the arbitrage-free interpolation and extrapolation. In section~\ref{numEx} some numerical experiments are presented as well as stability of the fitted parameters with time is analyzed. The last section discusses some remaining issues.

\section{Model} \label{model}
Before we describe our construction, it is interesting to note that traditional parametric models represent the smile as some polynomial function of $z$. One of the reasons for doing this is that according to \cite{Cont2002} the IV patterns across moneyness vary less in time than when expressed as a function of the strike. Also, there is an additional computational benefit by regressing at moneyness rather than at the strike prices, since the function is of a simpler form, and, therefore, the estimation algorithm converges faster.

A typical study is that of \cite{Alentorn04} where using data in the FTSE 100 index, the following models were tested:
\begin{eqnarray}  \label{classModel}
\sigma(z) &=& \beta_0 + \beta_1 z + \beta_2 z^2 + \epsilon \\
\sigma(z) &=& \beta_0 + \beta_1 z + \beta_2 z^2 + \beta_3 T + \beta_4 T z \epsilon, \nn
\end{eqnarray}
\noindent where $\beta_i, \ i\in [0,4]$ are the regression parameters that usually are a function of time, therefore \eqref{classModel} with the fixed coefficients represents just one term $T=const$ of the volatility surface. In \cite{Borovkova2008} the authors use a similar regression. They also noticed that the parabolic shape of the implied volatility function for a fixed maturity is the average shape of the actual volatility functions. Note that increasing the power of the polynomial volatility function (from two to three or higher) does not really offer a solution here, since this volatility function will still be the same for all maturities. Quadratic profile of the implied volatility as a function of $z$ is also supported by PCA analysis of the implied volatility surface (\cite{Cont2002, Alexander2001, Fengler2003}).

Comparison of these models with the market data showed that they are able to capture a form of the volatility smile in the ATM region while often fail at wings. Another problem is fitting the smile close to expiration. Here $T \rightarrow 0$ implies $z \rightarrow \infty$, and the volatility at wings tends to infinity which is not supported by the market data. Therefore, the regression coefficients $\beta_1, \beta_2$ must tend to zero, and the fitting function degenerates in this limit. This poses a real problem for the optimization routine (it never converges to such a limit).

Gatheral in his SVI parametrization uses another form
\begin{equation}  \label{gathModel}
w(\chi;a,b,\sigma,\rho,m) = a + b \left\{ \rho(\chi -m) + \sqrt{(\chi-m)^2 + \sigma^2} \right\}.
\end{equation}
Here $w$ is the total implied variance, $\chi = \log(K/F)$, $a$ gives the overall level of variance, $b$ gives the angle between the left and right asymptotes, $\sigma$ determines how smooth the vertex is, $\rho$ determines the orientation of the graph, and changing $m$ translates the graph.

This form of the implied volatility surface is motivated by an asymptotic no-arbitrage argument, pioneered by \cite{Hodges:96}, \cite{Gatheral1999} and later \cite{Lipton2001}
who mentions that the resulting $IV(\chi)$ bounds are $O(|\chi|^{1/2})$ for large $|\chi|$. This was then further extended by the familiar results of \cite{Lee2004} that the total implied variance should be linear in $\chi$ at wings $\chi \to \pm \infty$ with the slope $0 < \phi(\infty) < 2$. As applied to the SVI \cite{Gatheral2005}) derives necessary and sufficient conditions for the IV surface to be arbitrage free and shows how this parametrization fits the IV surfaces generated by various currently popular models, including the stochastic volatility and jump models. Also some examples are provided where the SVI well fits the actual IV surfaces - even the notoriously hard-to-fit very short expirations.

SVI updated with the arbitrage-free interpolation and extrapolation, \cite{GatheralJacquier2014} and the latest versions of the quadratic regressions mentioned above work well in many situations. In our experience, however, we would need another model which combine capabilities of the latter models with better flexibility. For instance, i) the model should be capable of fitting both smile and skew using the same regression (which could be a problem with the quadratic model); ii) it would be good to have a separate model parameter which determines location of the smile minimum and could be calibrated to the market data (e.g., in SVI this location is predetermined by the values of the model parameters); iii) the behavior at wings could be sublinear (see below) while in the original SVI it is strictly linear, etc.

From this prospective we do our construction of a new parametric model based on the following assumptions.
\begin{enumerate}
\item As the independent variables of the parametric regression we choose the normalized strike $z$ defined in \eqref{z} and time to maturity $T$.

\item In the form presented in \eqref{fit} the model is capable to simulating a different behavior of the smile at call and put wings (\cite{ZhaoHodges2013}). Such a situation could be helpful when modeling commodities where one wing could demonstrate a linear behavior while the other one - sublinear. For the sake of brevity, however, when doing an asymptotic analysis of the model we omit
a detailed discussion of sub linearity (to be discussed elsewhere), and concentrate at the case where the variance smile at wings is linear in $z$.

\item As there exist multiple justifications that the smile is not symmetric in the $z$ space, it is highly desirable to fit the call and put wings independently.

\item The parametric function must be continuous in $z$.

\item It should be well-behaved close to expiration.

\item We fit the term structure of the IV term-by-term, i.e., first the variance curve at the first maturity $T_1$, than the variance curve at the second maturity $T_2 > T_1$, etc. Therefore, we do not consider the dependence of the regression parameters on time. However, we do discuss how to build the whole arbitrage-free IV surface.

\item  As a possible extension of this approach one can rely on the definition of $z$ where the calendar clock $T$ is replaced with a business clock $T_v$. Here we just mention this opportunity which apparently improves the fitting capability of the model, especially close to expiration, but don't discuss it in detail.

\item The number of parameters must be minimal.

\item The parametric function must be fast to evaluate
\footnote{For instance, a recent extension of the SVI model proposed in \cite{ZhaoHodges2013} utilizes the Kummer hypergeometric functions which makes the computation expansive.}.

\item The whole IV surface should respect the no-arbitrage conditions.
\end{enumerate}

Given $T$, our new parametrization of one term at the IV surface reads
\begin{align}  \label{fit}
w(z) &= w_c + {\cal S}_C \dfrac{y}{1 + y^2} + F(y)\sqrt{T} \sum_{i=1}^n a_i Y^i(y) \\
y &= z-{\cal C}, \quad Y(y) \equiv \left\{
\begin{array}{ll}
\dfrac{1}{\alpha} \mathfrak{S}\left( - \alpha y \right) & y \le 0 \\ \\
\dfrac{1}{\beta} \mathfrak{S}\left( - \beta y \right), & y > 0
\end{array}
\right.
\nn
\end{align}
\noindent where $w(z)$ is the total implied variance, $w(z) = I^2(z) T, \ I(z) $ is the implied volatility, $n$ determines the maximum degree of the polynomial on $Y(y)$, and $\mathfrak{S}(x)$ belongs to the class of the so-called sigmoid functions, \cite{Seggem2007}. The sigmoid functions tend to some constant at both ends when the argument $x$ tends to $\pm \infty$, and vanish at $x=0$. Many natural processes, including those of complex system learning curves, exhibit a progression from small beginnings that accelerates and approaches a climax over time. Besides the logistic function, sigmoid functions include the ordinary arctangent, the hyperbolic tangent, the Gudermannian function, and the error function, but also the generalized logistic function and algebraic functions like $x/\sqrt{1+x^2}$.

In \eqref{fit} the function $F(y)$ defines the model behavior at wings. It could be chosen in such a way that close to $y=0$ we have $F(y) \propto |y|^{\alpha_0}$ while $F(y) \to y^{\alpha_+}, \ \to \infty$, and $F(y) \to (-y)^{\alpha_-}, \ y \to -\infty$ where $0 < \alpha_- \le 1, \ 0 < \alpha_+ \le 1, \ 0 < \alpha_0$ are some constants. This construction is accounting for both linear and sublinear behavior of the regression at wings. However, in this paper we will explore only the case $F(y) \equiv |y|$, so the sublinear case will be discussed elsewhere.

From the performance point of view, we want  $w(z)$ to be computed with the minimal possible number of computer operations. This guides us in choosing $\mathfrak{S}(x) = \mbox{erf}(x)$ due to the approximation
\[
\mbox{erf}(x) \approx 1 - (1 + a_1 x + a_2 x^2 + ... + a_6 x^6)^{-16},
\]
\noindent with the maximum error $3\cdot 10^{-7}$, where $
 a_1 = 0.0705230784, a_2 = 0.0422820123, a_3 = 0.0092705272, a_4 = 0.0001520143,
 a_5 = 0.0002765672, a_6 = 0.0000430638$. This approximation is valid for $x \ge 0$. To use it for the negative $x$, exploit the fact that $\mbox{erf}(x)$ is an odd function, so $\mbox{erf}(x) = -\mbox{erf}(-x)$ (\cite{as64}).

It is worth mentioning that using polynomial functions in $\arctan(z)$ was a popular choice among practitioners a while ago, however we don't put this restriction. Also for clarity we fix $n=2$ and provide a special notation for $\mathcal{S} \equiv a_1, \ \mathcal{K} \equiv a_2$. The reason for this notation will become clear right below.

Under these assumptions, $w(z)$ in \eqref{fit} has 7 parameters:
\bi
\item ${\cal C}$ - shift. This is an edge point between the left and right branches of the smile. For equity options ${\cal C}\approx 0$, i.e. this is close to the ATM point. Then, the left branch is a put wing while the right branch is a call wing. For index options the minimum of the smile is usually shifted into positive $z$. The parameter ${\cal C}$ just reflects the value of this shift. Note, that the smile is $C_2$ at $z={\cal C}$. Indeed, the direct differentiation of $Y(z)$ in the Eq.~(\ref{fit}) shows that the first derivative is continuous and reads $Y'(z)\left|_{z={\cal C}} = -1 \right.$, while the second derivative vanishes.

\item $w_C$  - this is the variance at $z = {\cal C}$.

\item ${\cal S_C}$ - this parameter determines skew of the smile at $z={\cal C}$.

\item $\alpha$ - this is a put wing parameter which determines how steep the put wing should be.

\item $\beta$ - this is a call wing parameter which determines how steep the call wing should be.

\item ${\cal S}$ - this parameter determines skew of the smile outside of the region $0 \le z \le \cal C$.

\item ${\cal K}$ - this parameter determines kurtosis of the smile outside of the region $0 \le z \le \cal C$.
\ei
In the limit $\alpha \rightarrow 0$ or $\beta \rightarrow 0$ we obtain $Y(z) \rightarrow  {\cal C}-z$.

Further on, for getting better results we need a minor refinement of the model. As the fitted variance $w(z)$ is expected to be at least $C_2$ continuous in $z$, it would be better to eliminate such a non-continuous function as $|z- {\cal C}|$. This could be relatively easy done if we find a continuous approximation of the function $|z-{\cal C}|$. Among various possible functions we chose that
\begin{equation}\label{approx}
    |y| \approx y \tanh [p y],
\end{equation}

\noindent  where $p$ is some constant parameter. Choosing $p$ big enough, say 1000, gives us highly accurate approximation of $|y|$ which is infinitely continuous.

\section{Asymptotic analysis and meaning of the parameters}
Below we provide an asymptotic analysis of the model to reveal the financial meaning of all the model parameters.

\subsection{Behavior at $z = \mathcal{C}$}
To better understand why one needs a linear correction term, consider the asymptotic behavior of the smile. As $z \rightarrow {\cal C}$ the function $w(z)$ behaves like
\begin{equation}  \label{asym_sh}
w(z) \approx w_C  + \mathcal{S_C} y - \left(\mathcal{S_C} + p \mathcal{S} \sqrt{T}Y_y(0)\right)y^3  + O\left( y^4\right).
\end{equation}

Thus, this is a polynomial function of $z-{\cal C}$ which is similar to what \cite{Dumas98} model does. More rigorously, it is linear in $y$ at small $y$ if $p < 1/y$, and quadratic if we choose $p \approx 1/y$ at small $y$. Also as the parameter $\alpha$ determines the steepness of the smile in the put wing, it is reasonable to have an independent parameter to better shape the linear part of the smile near $z=\cal C$. That is why in the Eq.~(\ref{fit}) we introduced an extra term which is proportional to $z {\cal S_C}$ at $z \approx \cal C$, and vanishes at $z \rightarrow \infty$.

From \eqref{asym_sh} it is clear that $w_C$ is the total variance at $z=\mathcal{C}$, and $\mathcal{S_C}$ is the skew at $z=\mathcal{C}$, while the kurtosis at $z=\mathcal{C}$ vanishes. Also, it is seen that varying $p$ one can change the value of higher moments, which, however, for our analysis is not that important.

Thus, we can interpret the coefficients $w_{C}, \mathcal{S}_C$ and $\mathcal{C}$ as some form of adjustment for the critical point not being at $z=0$.

Note that since the derivatives bear no dependence on $\beta$ (or $\alpha$), the model is indefinitely continuous around $z=C$.

\subsection{Behavior ATM}
Let's consider the behavior of our model at the money when the strike $K$ is equal to the forward price $F$, and so $z = 0$. To simplify the analysis, we assume $\mathcal{C} > 0, \ p\mathcal{C} \gg 1$. The value of $p$ could be always chosen such that $p\mathcal{C} \gg 1$ unless $\mathcal{C} = 0$. Then $\tanh(p\mathcal{C}) \approx 1$ if $\mathcal{C} > 0$, and $\tanh(p\mathcal{C}) \approx -1$ if $\mathcal{C} < 0$.  For easy of notation, denote $A \equiv Y(-\mathcal{C}), \ A' \equiv Y_y(-\mathcal{C}), \
A'' \equiv Y_{yy}(-\mathcal{C})$.

From the \eqref{fit} the ATM variance is given by
\begin{equation}\label{ATMvar}
    w_0 = w_{C} - \frac{\mathcal{C}}{1+\mathcal{C}^2}\mathcal{S_C} + A \mathcal{C} \sqrt{T} \tanh(p \mathcal{C}) \frac{\mathcal{S} \alpha + A \mathcal{K}}{\alpha^2} + O(y+\mathcal{C}).
\end{equation}
Accordingly, the ATM skew is approximately given by
\begin{equation} \label{ATMskew}
\mathbb{S}_{\mathrm{ATM}} = -\frac{\left(\mathcal{C}^2-1\right) }{\left(\mathcal{C}^2+1\right)^2} S_\mathcal{C}
+ \tanh(p \mathcal{C}) \sqrt{T}
 \left[ -A (\mathcal{S} + \mathcal{K} A) + \mathcal{C} A' (\mathcal{S} + 2 \mathcal{K} A) \right] + O(y+\mathcal{C}),
\end{equation}
\noindent and the ATM kurtosis is
\begin{align} \label{ATMkurt}
\mathbb{K}_{\mathrm{ATM}} &= -2 S_\mathcal{C}\frac{\mathcal{C} \left(\mathcal{C}^2-3\right) }{\left(\mathcal{C}^2+1\right)^3} \\
&+ \sqrt{T} \tanh(p \mathcal{C})
\left[ -2 A' \left( \mathcal{K} (2 A - \mathcal{C} A') + \mathcal{S} \right)
+ \mathcal{C} A'' \left( 2 A \mathcal{K} + \mathcal{S} \right) \right]
+ O(y+\mathcal{C}). \nonumber
\end{align}

Further on, we want to determine a connection between the inflection point $\mathcal{C}$ and parameters of the smile ATM. In order to do that, first suppose $\mathcal{C}$ is small, but our assumption $p\mathcal{C} \gg 1$ is still preserved because of a big $p$. Also, in our numerous experiments where we calibrated this model to various equity and index options with a wide range of maturities and strikes it was observed that a typical value of $\mathcal{K}$ is about 1.0, $\mathcal{S}$ is of the order of 1.0, and $\alpha$ varies from 0 to 5. Therefore, from \eqref{ATMskew} we obtain
\begin{equation} \label{solution}
C^2 =  \frac{\mathcal{S}_C - \mathbb{S}_{\mathrm{ATM}} }{3 \mathcal{S}_C - 2 p \mathcal{S}\sqrt{T} Y_y(0)}
\end{equation}

Thus, our assumption that the value of $\mathcal{C}$ is small is true if $\mathcal{S_C} - \mathbb{S}_{\mathrm{ATM}} \ll {3 \mathcal{S}_c - 2 p \mathcal{S}\sqrt{T} Y_y(0)}$. As $\mathcal{C}$ is small (in other words, close to the ATM) the difference $\mathcal{S_C} - \mathbb{S}_{\mathrm{ATM}}$ also has to be small.

At very small $T$ the above solution transforms to
\begin{equation} \label{ATMskew2}
\mathbb{S}_{\mathrm{ATM}} = -\frac{\left(\mathcal{C}^2-1\right) }{\left(\mathcal{C}^2+1\right)^2} S_\mathcal{C}
\end{equation}

At small $\mathcal{C}$ this equation has the root
\[
\mathcal{C} = \sqrt{\frac{\mathcal{S_C} - \mathbb{S}_{\mathrm{ATM}}}{3 \mathcal{S_C}}}
\]

If $\mathcal{C}$ is positive, the ATM point belongs to the put wing, and $\mathbb{S}_{\mathrm{ATM}} < 0$. Therefore, $\mathcal{S_C}$ has to be positive in order for the Eq.~(\ref{solution}) to be consistent.

Note, that the Eq.~(\ref{solution}) does not contain $\alpha$ or $\beta$, therefore it is valid regardless of whether $\mathcal{C}$ is positive or negative. Also from the Eq.(\ref{ATMskew}) it follows that the minimum of the smile at $\mathcal{C}=0$ does not coincide with the ATM point.

\begin{figure}[H]
\begin{minipage}[b]{0.4\textwidth}
\begin{center}
\fbox{\includegraphics[height=3in, width = 3in]{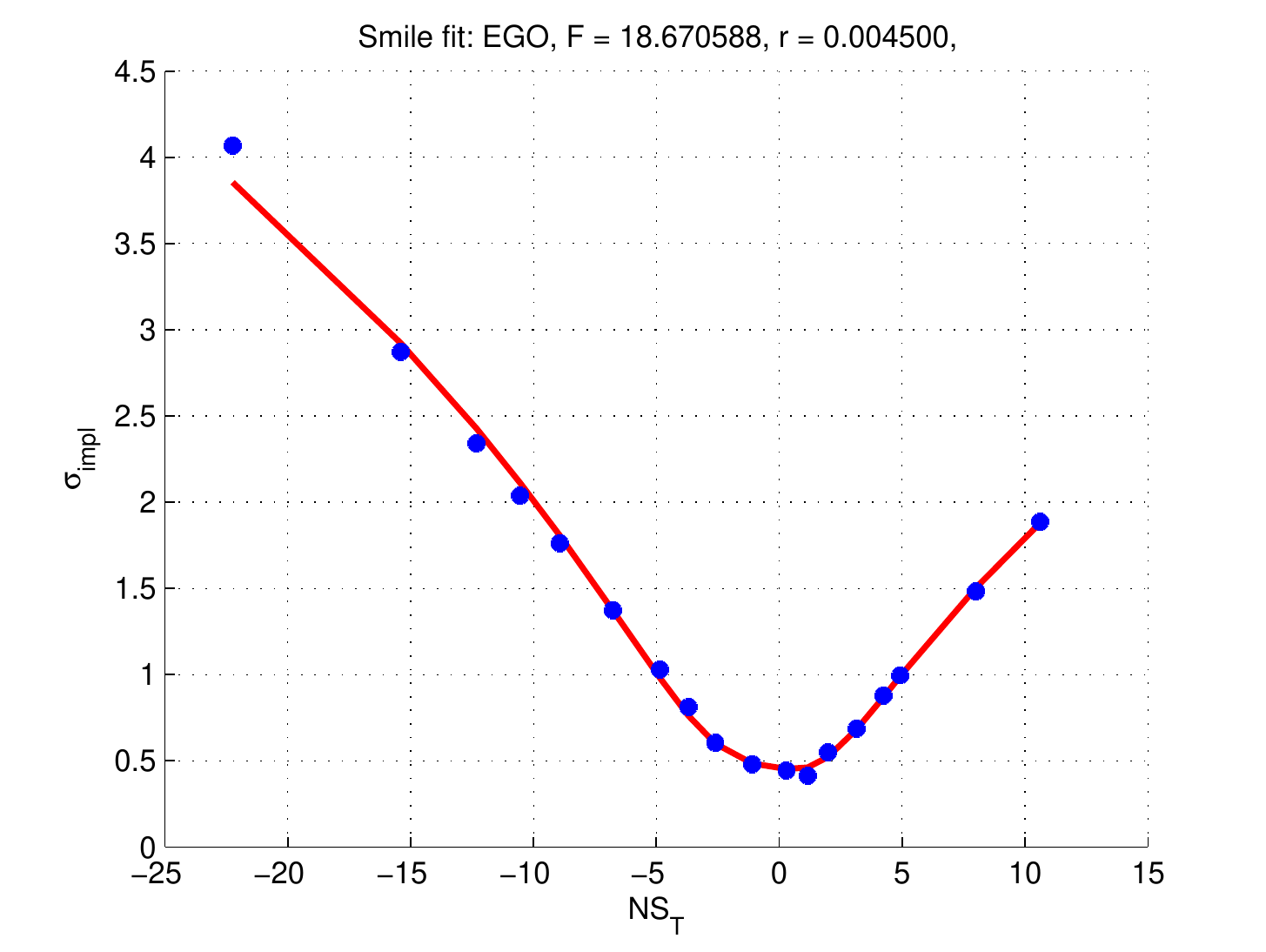}}
\caption{Fitting of the IV smile for EGO, $T=$10/15/2010}
\label{FigEGO}
\end{center}
\end{minipage}
\hspace{0.1\textwidth}
\begin{minipage}[b]{0.4\textwidth}
\begin{center}
\fbox{\includegraphics[height=3in, width=3in]{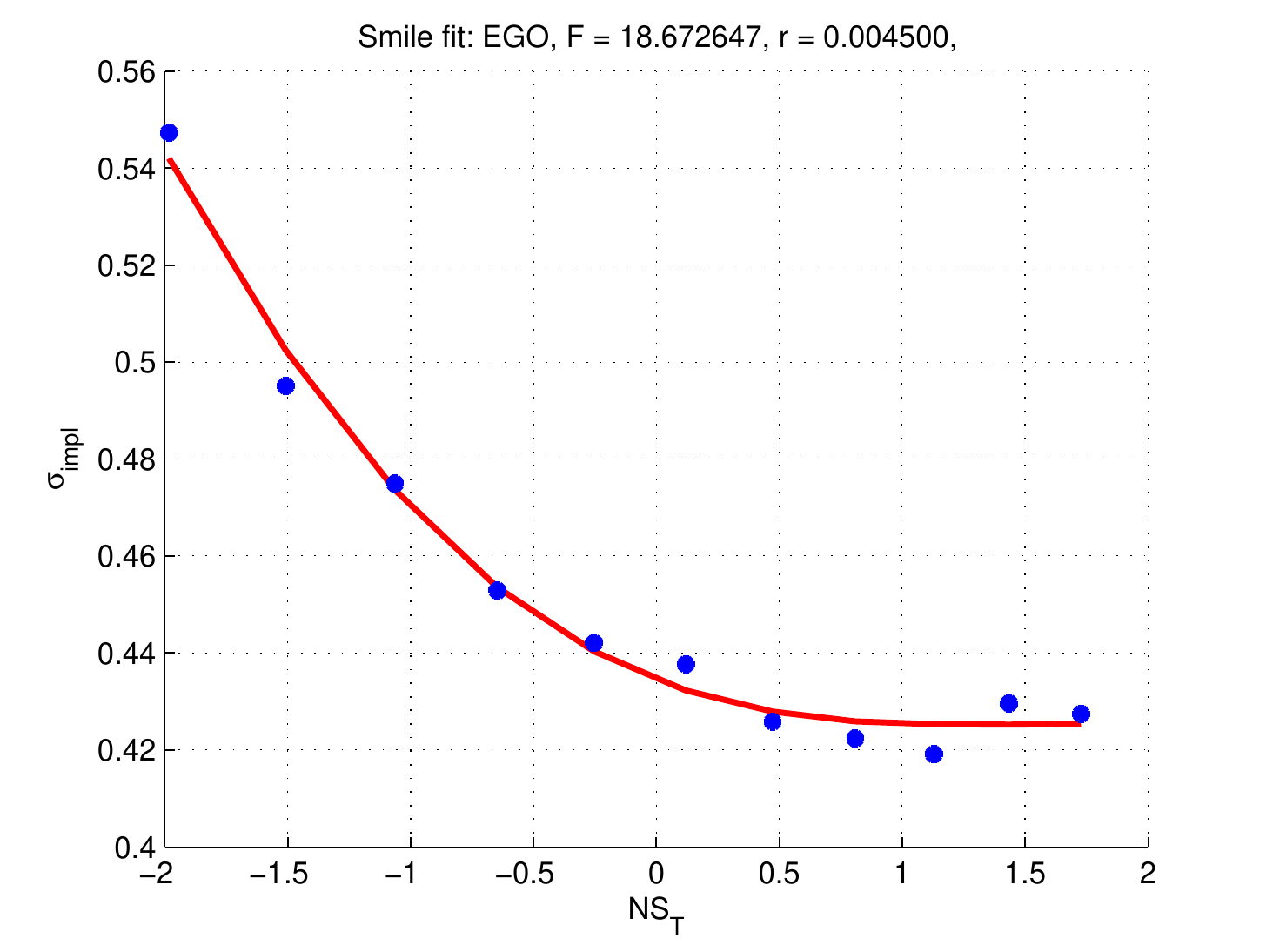}}
\caption{Fitting of the IV smile for EGO, $T=$11/19/2010.}
\label{FigEGO1}
\end{center}
\end{minipage}
\end{figure}

To illustrate this analysis, here we provide an example of a real smile computed using the proposed model. We run this test on Oct. 7, 2010 and fit the implied volatility of options written on the Eldorado Gold Corporation (EGO) stock with expire on Oct. 15, 2010. The results of fitting are given in Fig.~\ref{FigEGO} where $NS_t \equiv z/\sigma_{ATM}$.

Parameters of the fit found by calibration are given in Tab.~\ref{Tab1}:
\begin{table}[H]
\begin{center}
\begin{tabular}{|c|c|c|c|c|c|c|}
\hline
$w_{C}$ & $\mathcal{S}_C$ & $\mathcal{C}$ & $\mathcal{S}$ & $\mathcal{K}$ & $\alpha$ & $\beta$ \\
\hline
0.1652 & -0.04302 & 0 & -0.20 & 1.035 & -0.42623 & 0.60308 \\
\hline
\end{tabular}
\caption{Experiment 1, parameters of the fit.}
\label{Tab1}
\end{center}
\end{table}

Thus, this smile does not demonstrate any shift of the minimum from ATM.

In the second example we fitted the next term of the same product. The results are given in Fig.~\ref{FigEGO1}. Parameters of the fit found by calibration are given in Tab.~\ref{Tab2}:

\begin{table}[H]
\begin{center}
\begin{tabular}{|c|c|c|c|c|c|c|}
\hline
$w_{C}$ & $\mathcal{S}_C$ & $\mathcal{C}$ & $\mathcal{S}$ & $\mathcal{K}$ & $\alpha$ & $\beta$ \\
\hline
0.16775 & 0 & 0.5769 & -0.003 & 0.11 & -0.0004 & 2.7457 \\
\hline
\end{tabular}
\caption{Experiment 2, parameters of the fit.}
\label{Tab2}
\end{center}
\end{table}

It is seen that in the test $\mathcal{C}$ is not a small parameter, therefore simple approximations suggested in the above cannot be used in this case.

\begin{figure}[H]
\begin{center}
\fbox{\includegraphics[height=3in, width=3in]{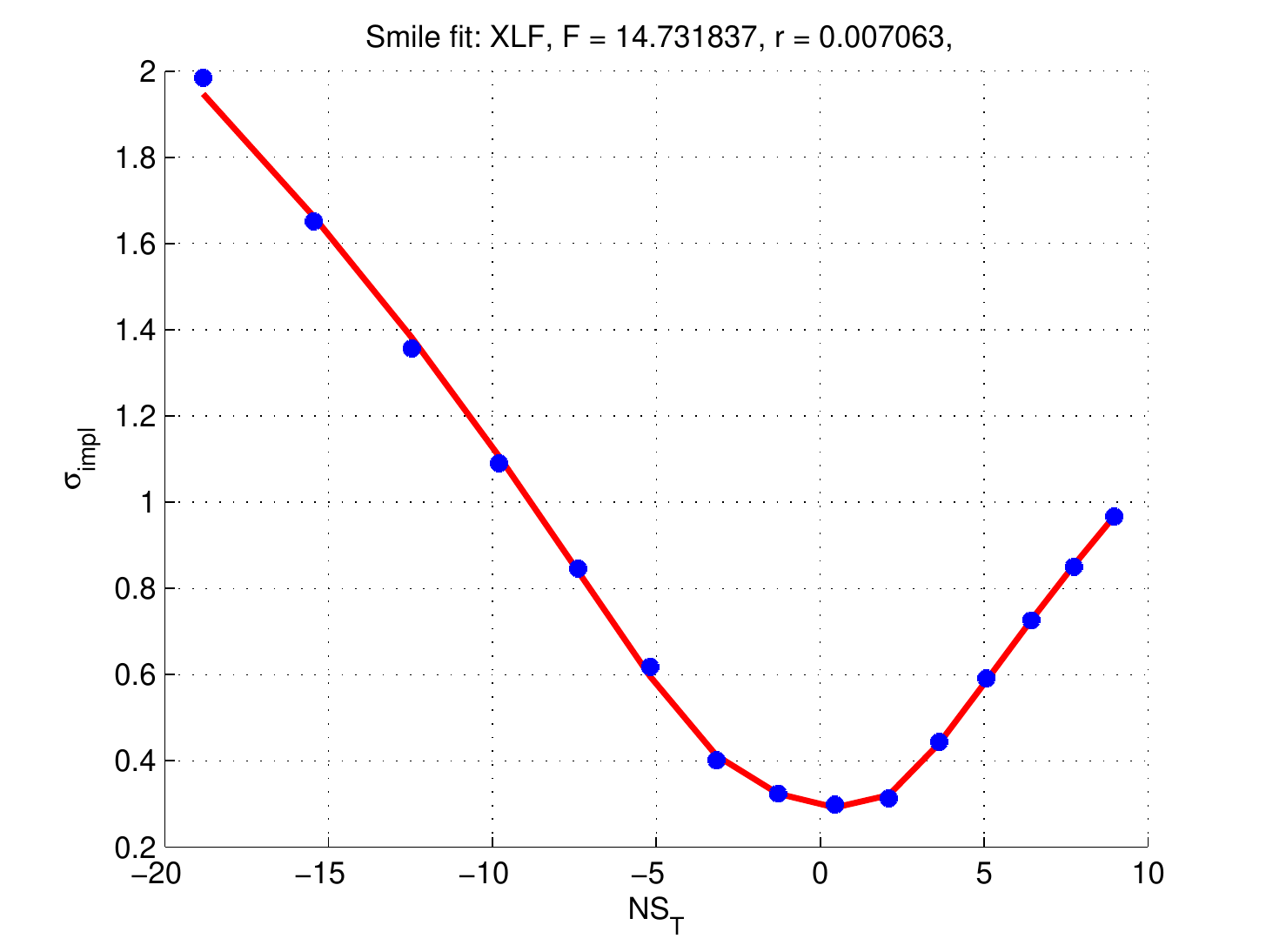}}
\caption{Fitting of the IV smile for EGO, $T=$10/15/2010.}
\label{FigXLF}
 \end{center}
\end{figure}

The third example is given for options written on Financial Select Sector SPDR Fund (XLF) stock, also for the front term Oct.15, 2010. The results are given in Fig.~\ref{FigXLF}.

Parameters of the fit found by calibration are given in Tab.~\ref{Tab3}:
\begin{table}[H]
\begin{center}
\begin{tabular}{|c|c|c|c|c|c|c|}
\hline
$w_{C}$ & $\mathcal{S}_C$ & $\mathcal{C}$ & $\mathcal{S}$ & $\mathcal{K}$ & $\alpha$ & $\beta$ \\
\hline
0.0703 & -0.038 & 0.0032 & -0.2 & 0.99 & 0.5741 & -0.80175 \\
\hline
\end{tabular}
\caption{Experiment 3, parameters of the fit.}
\label{Tab3}
\end{center}
\end{table}

In this test the calibrated value of $\mathcal{C}$ is small, so one can use the proposed approximations which connect the ATM skew and kurtosis with the value of $\mathcal{C}$.

\subsection{Behavior At Infinity}
As $K \to \infty$ at fixed $S$, so does $z$. Assume, for example, that $\mathfrak{S}(x) \equiv \arctan(x)$. Expanding variance in series around positive infinity, we have
\begin{equation} \label{plInf}
w(z) \approx
w_{C} + \frac{\sqrt{T}}{\beta^3}\left(\beta \mathcal{S} - \pi \mathcal{K}\right)
+ \pi  \sqrt{T} \frac{\frac{1}{2} \pi \mathcal{K} - S \beta}{2\beta^2} y
 + O \left(1/y\right).
\end{equation}

Hence, the variance is linear in log-moneyness $\chi$ at positive infinity, with the slope
\begin{equation} \label{extInf}
\phi(\infty) = \frac{1}{2} \pi \beta^{-2} (\frac{1}{2} \pi \mathcal{K} - \mathcal{S} \beta).
\end{equation}
This well agrees with the result of \cite{Lee2004}. Thus, our interpretation of $\beta$ follows: this parameter controls the slope of the smile at the infinite strike.

At $K \to 0, \ z \to -\infty$. Expanding variance in series around negative infinity, we have
\begin{equation} \label{plInf2}
w(x) \approx w_{C}
- \frac{\sqrt{T}}{\alpha^2} \left(\alpha\mathcal{S} + \pi \mathcal{K}\right)
- \pi \sqrt{T} \frac{\frac{1}{2}\pi \mathcal{K} + \mathcal{S} \alpha}{2\alpha^2} y
+ O \left(1/y\right).
\end{equation}

Hence, the variance is also linear in log-moneyness $\chi$ at negative infinity, with the slope
\begin{equation} \label{extMinInf}
 \phi(-\infty) =  - \frac{1}{2} \pi \alpha^{-2} (\frac{1}{2} \pi \mathcal{K} + \mathcal{S} \alpha).
 \end{equation}
This also agrees with the result by \cite{Lee2004}. Accordingly, our interpretation of $\alpha$ is: this parameter controls the slope at strike close to zero.

Close to expiration $z$ tends to infinity. However, for our function in the Eq.~(\ref{fit}) this is not a problem. Indeed, at $T \rightarrow 0, z \rightarrow \infty$, the product $z \sqrt{T} \rightarrow \log{K/F}$, therefore from \eqref{plInf}
\begin{equation}  \label{asym_T}
w(z) \rightarrow w_C + \dfrac{\pi}{2\beta^2} \left(\frac{1}{2}{\cal K} \pi - \beta {\cal S}\right) \log \dfrac{K}{F},
\end{equation}
\noindent As mentioned in \cite{Medvedev2008, Ledoit2002} in diffusion models, the ATM implied volatility is known to converge to the spot volatility when $T$ goes to zero. In our case from \eqref{asym_T} the ATM value $w(z)|_{z=0} = w_C$, which implies $I(z) = \sigma_C = const, \ w_C = \sigma_C^2 T$. This provides another interpretation of the parameter $\sigma_C$ as the IV of the underlying stock at $T=0$.

\section{No arbitrage conditions} \label{na}

In contrast to the case where the IV surface is built based on some model, e.g., a stochastic volatility model, which guarantees no-arbitrage by construction, using regressions doesn't provide such a nice feature {\it per se}. Therefore, in the latter cases a special care should be taken under calibration in order not to introduce arbitrage into the IV surface. See, for instance, \cite{AndreasenHuge2010, LiptonSepp2011iv,GatheralJacquier2014} and references therein.

The no-arbitrage conditions could be expressed in various forms. One of the approaches is to say that the local volatility function must be non-negative. The reason for that is that the local volatility function is directly related to the pdf (density) of the underlying, which in turn has to be non-negative. Then using Dupire's formula for the local volatility as this is done in \cite{Gatheral2006}, i.e., representing it via the $w(z)$ function, one gets
 \begin{equation} \label{lv}
\sigma_{loc}^2(T, K) = \frac{\dd_T w}{\left(1-\frac{\chi\dd_\chi w}{2 w}\right)^2
- \frac{(\dd_\chi w)^2}{4}\left(\frac{1}{w}+\frac{1}{4}\right)+\frac{\dd^2_\chi w}{2}},
\end{equation}
The nominator of this expression is the so-called calendar spread, and the denominator of it is equivalent to the so-called butterfly spread (which for the call option with price $C(T,K)$ is defined as $\sop{C(T,K)}{K}$,  \cite{GatheralJacquier2014}). Both spreads must be non-negative for no-arbitrage.

However, as shown in \cite{CarrMadan2005}, one more condition is required in addition to the above mentioned, which tells that so-called vertical call spread (which for the call option with price $C(T,K)$ is defined as $\fp{C(T,K)}{K}$) should be negative for the call options, or the vertical put spread should be positive for the put options. For the IV these conditions for the vertical spreads could be transformed to the following,  \cite{Carr2004}
\begin{equation} \label{pVS}
\dfrac{R(d_2)}{\sqrt{T}} \le K \fp{I(K,T)}{K} \le \dfrac{R(-d_2)}{\sqrt{T}},
\end{equation}
\noindent where $ R(d) \equiv \dfrac{1-N(d)}{N'(d)}$ is Mill's ratio, $N(d)$ is the normal cdf, and $d_2$ comes from the Black-Scholes formula. The convenience of such a representation lies in the fact that Mill's ratio for the standard normal distribution reads
\[ R(x) = e^{x^2/2} \sqrt{\frac{\pi}{2}}\mbox{erfc}\left(\frac{x}{\sqrt{2}}\right). \]
The latter can be efficiently computed by the particularly simple continued fraction representation at $ x > 1$
\begin{equation}
R(x) = \cfrac{1}{x
          + \cfrac{1}{x + \cfrac{2}{x + ...} } },
\end{equation}
\noindent or by using Taylor series expansion at $ 0 \le x \le 1$, \cite{MillsRatio2014}.
The IV surface should also satisfy the asymptotic conditions discussed in the previous section, namely: the slope $\phi(\infty)$ of the call wing at $z \to \infty$ should be $ 0 \le \phi(\infty) \le 2$, and the slope of the put wing $\phi(-\infty)$ at $z \to -\infty$ should be $ 0 \ge \phi(-\infty) \ge -2$.

Being equipped with all these no-arbitrage conditions, the next step to consider is the construction of the IV surface in the domain $(T,K)$. This is not a problem if, say, we want to have the IV surface to be defined at some discrete grid in the $(T,K)$ space (that could be a grid where we want the local volatility function to be determined - a standard approach when one calibrates the LSV model to the market data):
$\mathbb{G} : [T_i \times K_j], \ i \in [1,N], \ j \in [1,M]$ under two assumptions made: i) for every grid node $(i,j)$ there exists a market quote $Q(T_i,K_j)$ which is an option price (call or put, or both); ii) there is no need to ever know the IV at other possible values of $T,K$ which don't belong to $\mathbb{G}$. Certainly, in practice both assumptions are unrealistic. Therefore, some kind of interpolation/extrapolation which preserves no-arbitrage is necessary.

Therefore, in order to calibrate our model to the market data such that not every node on the computational grid is provided with a corresponding market quote, a special calibration algorithm ws elaborated on which in more detail is described in Appendix~\ref{ap}.

\subsection{No-arbitrage interpolation on the grid}

Various approaches were discussed in the literature with regard to this problem. For instance, an arbitrage-free interpolation was considered in \cite{AndreasenHuge2010, Fengler2005, GatheralJacquier2014} (see also references therein). Here, however, we suggest another approach, which is similar in spirit to that in \cite{GatheralJacquier2014}.

If one works in the $z$ space given $T$ the usual approach would be to choose some number $\gamma$ such that all $z = [z_1...z_M]$ for this term are in the range $-\gamma < z/\sigma_* < \gamma$. Here $\sigma_*$ is some normalization constant which doesn't depend on $T$. By financial meaning, $\sigma_*$ could be chosen as the ATM IV which corresponds to the shortest maturity, This is the most liquid strike of the instrument, and usually it is pretty well-known from the market data. In other words, for the IV surface just a range of $\gamma$ standard deviations in both up and down directions from the ATM is taken into account. Outside of this domain the remaining strikes are treated to be illiquid, and, therefore, they are taken out of consideration. In practical applications $\gamma = 5$ could be chosen, but this assumption could be easily relaxed.

Another situation is if we want the IV surface to be a building block of the numerical method which solves the pricing/calibration problem using the local stochastic volatility model. The idea is first to calibrate the IV surface to the market quotes of the vanilla options, and then compute the local volatility surface using the Dupire's formula. In this case, we need the values of the local volatility function not only at strikes and maturities available were the market data are available, but at all nodes in the computational domain $K,T$ which is these calculations. In more detail, in this case we fist define a fixed domain in $K$ space: $[K_1...K_M]$. Accordingly, for the $z$ variable we have a map $z_i = \log (K_i/F(T_j))/\sqrt(T_j)$ which depends on the current expiration $T_j, \ j=1...N$. In other words, we work on the $\mathbb{G}$ grid which was described in the above.

Provided by a set of the IV market data for expirations $T_1 < T_2 < ... < T_m$ (these expirations in general don't coincide with the temporal nodes of the grid $\mathbb{G}$, but could be a subset of that) we calibrate our model term by term based on the algorithm of Appendix~\ref{ap}. To remind, this algorithm takes into account the entire set of the no-arbitrage constraints at every point on the given grid $\mathbb{G}$. By construction, despite the market provides the option quotes per strikes, the grid was built in the $z$ space, not in the strike $K$ space. At the end we obtain all values $w(z,t)$ where $z \in [-\gamma \sigma_* = z_1, ..., z_N = \gamma \sigma_*], T \in [T_1, ..., T_m]$. Given thus found $w(z,t)$ the corresponding undiscounted call and put prices can be further obtained by using the Black-Scholes formula afterwards.

After this step is completed the arbitrage-free values $w(z,t)$ become available for the terms with expirations $T_1 < T_2 < ... < T_m$. For the sake of clearness let us denote them as $w_1(z,t) \equiv w(z,t), \ t \in [T_1,...,T_m]$. However, our grid $\mathbb{G}$ by construction also might contain some other expirations $\hat{T}_1,...,\hat{T}_l$ where $l$ is the total number of temporal nodes on the grid. Let us denote this set as $w_2(z,t)$. Also let us emphasize that the space nodes $z$ are the same for both $w_1(x,t)$ and $w_2(t)$ by construction.

Therefore, to find $w_2(z,t)$ next we need to interpolate $w_1(x,t)$ to the expirations of $w_2(z,t)$ at every point $z$ on the $\mathbb{G}$ grid. When doing that it is more convenient to proceed in the pricing space despite this is a bit more computationally intensive as we need to convert the IVs force to the prices at the beginning of this step, and back to the IVs at the end of this step. For the fitted terms after calibration is done we already know all parameters of the fit, so we are able to compute the call option value at any point at the $\mathbb{G}$ grid. And the no-arbitrage conditions were already respected in these points as well. Also we know the corresponding map $K_i \rightarrow F(T_j) e^{z_i \sqrt{T_j}}, i = 1...M, j = 1...N$.

Accordingly, in $K$ space the no-arbitrage conditions for the call option: non-negativity of the calendar and butterfly spreads and non-positivity of the vertical spread read
\begin{equation} \label{noarb}
\fp{C(K,T)}{T} \ge 0, \qquad \fp{C(K,T)}{K}  \le 0, \qquad \sop{C(K,T)}{K} \ge 0.
\end{equation}
Now chose a monotonic time interpolation of $C(K,T)$ at $K$=const of the form
\begin{equation} \label{interp}
C(K,T) = \alpha(T) C(K,T_1) + [1-\alpha(T)] C(K,T_2)
\end{equation}
\noindent where $T_1 < T < T_2$ and
\begin{equation} \label{alphaT}
\alpha(T) = \frac{a(T_2) - a(T)}{a(T_2) - a(T_1)},
\end{equation}
\noindent where $a(T)$ is some monotonic function. Obviously, $\alpha(T) \in [0,1]$, and $\alpha(T)$ doesn't depend on $K$. And this is a valid interpolation formula in a sense that the values of $C(K,T)$ at $T=T_1$ and $T=T_2$ coincide with $C(K,T_1)$ and $C(K,T_2)$. Also thus defined $C(K,T)$ provides
\begin{align} \label{calSpr}
\fp{C(K,T)}{T} &= \alpha(T) \fp{C(K,T_1)}{T} + [1-\alpha(T)] \fp{C(K,T_2)}{T} \\
&+ \fp{\alpha(T)}{T}[C(K,T_1) - C(K,T_2)]  \ge 0 \nonumber
\end{align}
\noindent if $\partial_T \alpha(T) < 0$, i.e., $\partial_T a(T) < 0$. That is because we constructed $C(K,T_1), C(K,T_2)$ such that they obey the no-arbitrage condition $\partial_T C(K,T)|_{T=T_i} > 0, \ i=1,2$.

It is easy to see that \eqref{interp} also solves the second and third lines in \eqref{noarb} provided that these conditions were met at $T=T_1$ and $T=T_2$. The latter follows from our construction at the previous step of the algorithm. Also it can be shown that this expression still preserves the extreme slopes of the interpolated terms (that are at $z \to \infty$ and at $z \to -\infty$) to follow the asymptotic conditions provided by \cite{Lee2004}. To see that, note that the latter could be represented in the form $C(K,T,I) < C(K,T,\sqrt{2 |\chi|/T})$. Therefore, our interpolation provides
\begin{align} \label{interpExt}
C(K,T,I) &= \alpha(T) C(K,T_1, I_1(K,T_1)) + (1-\alpha(T) C(K,T_2,I(K,T_2)) \\
&< \alpha(T) C(K,T_1, \sqrt{2 |\chi_1|/T_1}) + (1-\alpha(T) C(K,T_2,\sqrt{2 |\chi_2|/T_2})
\nonumber \\
&< C(K,T, \sqrt{2 |\chi|/T}). \nonumber
\end{align}
The last equality holds because we interpolate at $K$=const, $S$=const, so $C(K,T, \sqrt{2 |\chi|/T})$ is a function of $T$ only, and this is a concave function of $T$.

As far as extrapolation is concerned, in addition to the no-arbitrage conditions we need to prove that the extreme slopes of the extrapolated terms (that are at $z \to \infty$ and at $z \to -\infty$) still preserve the asymptotic conditions provided by \cite{Lee2004}.

We show that an extrapolation formula
\begin{equation} \label{extrap}
T^k C(K,T) = \alpha(T) T^k_1 C(K,T_1) + [1-\alpha(T)] T^k_2 C(K,T_2)
\end{equation}
\noindent with $k \in Re, \ k \le -0.5$ is suitable for this purpose.

Indeed, similar to \eqref{interp}
\begin{align} \label{interpExt1}
T^k C(K,T,I) &= \alpha(T) T^k_1 C(K,T_1, I_1(K,T_1)) + (1-\alpha(T) T^k_2 C(K,T_2,I(K,T_2)) \\
&< \alpha(T) T^k_1 C(K,T_1, \sqrt{2 |\chi_1|/T_1}) + (1-\alpha(T) T^k_2 C(K,T_2,\sqrt{2 |\chi_2|/T_2})
\nonumber \\
&< T^k C(K,T, \sqrt{2 |\chi|/T}). \nonumber
\end{align}
The last inequality holds because we interpolate at $K$=const, $S$=const, so
\begin{equation} \label{ff}
f(T) \equiv T^k C(K,T, \sqrt{2 |\chi|/T})
\end{equation}
 \noindent is a function of $T$ only, and $k < 0$ is chosen such that $f(T)$ is a convex function. The latter condition depends on the value of the interest rate $r$, and $\chi$. Usually, $k=-1$ is sufficient even for the ATM strikes \footnote{As it could be easily seen from analysis of the Black-Scholes formula for the call option prices this is the most sensitive region. Therefore, the choice of, e.g., $k=-0.5$ could make $f(T)$ to be concave close to ATM.}.
Also thus defined $C(K,T)$ provides
\begin{align} \label{calSpr1}
T^k \fp{C(K,T)}{T} &= \alpha(T) T^k_1 \fp{C(K,T_1)}{T} + [1-\alpha(T)] T^k_2 \fp{C(K,T_2)}{T} \\
&+ \fp{\alpha(T)}{T}[T^k_1 C(K,T_1) - T^k_2 C(K,T_2)] - k T^{k-1} C(K,T) \ge 0 \nonumber
\end{align}
\noindent if $\partial_T \alpha(T) < 0$. That is because we constructed $C(K,T_1), C(K,T_2)$ such that they obey the no-arbitrage condition $\partial_T C(K,T)|_{T=T_i} > 0, \ i=1,2$, and $T_1^k C(K,T_1) - T^k_2 C(K,T_2) < 0$ since $T_1 < T_2$ and $k = -1$. Thus, the calendar spread is non-negative for the call option. The other two no-arbitrage conditions obviously follow.

\section{Numerical experiments} \label{numEx}

\subsection{Stability of the fitted parameters}
In a typical experiment a volatility smile of an option written on the S\%P 500 index (SPX) was fitted using the proposed model. The raw data are collected for $T = 0.6247$ (228 days to expiration), $F = 76.58, T_v = 0.6215$. We find that our model fits the data pretty well, with parameters of the fit obtained by running the above described minimization algorithm, given in Tab.~\ref{Tab11}.
\begin{table}[H]
\centering
\begin{tabular}{|c|c|c|c|c|c|c|}
\hline
$w_C$ & ${\cal S}$ & ${\cal K}$ & $\alpha$ & $\beta$ & ${\cal C}$ & ${\cal S_C}$ \\
\hline
0.0435 & -0.763 & 74.5 & 3.12468 & 1.5000  & 0.2739 & -0.05921 \\
\hline
\end{tabular}
\caption {Values of the parameters obtained in the test}
\label {Tab11}
\end{table}

\begin{figure}[H]
\begin{minipage}[b]{0.4\textwidth}
\begin{center}
\fbox{\includegraphics[width=3in, height = 3in]{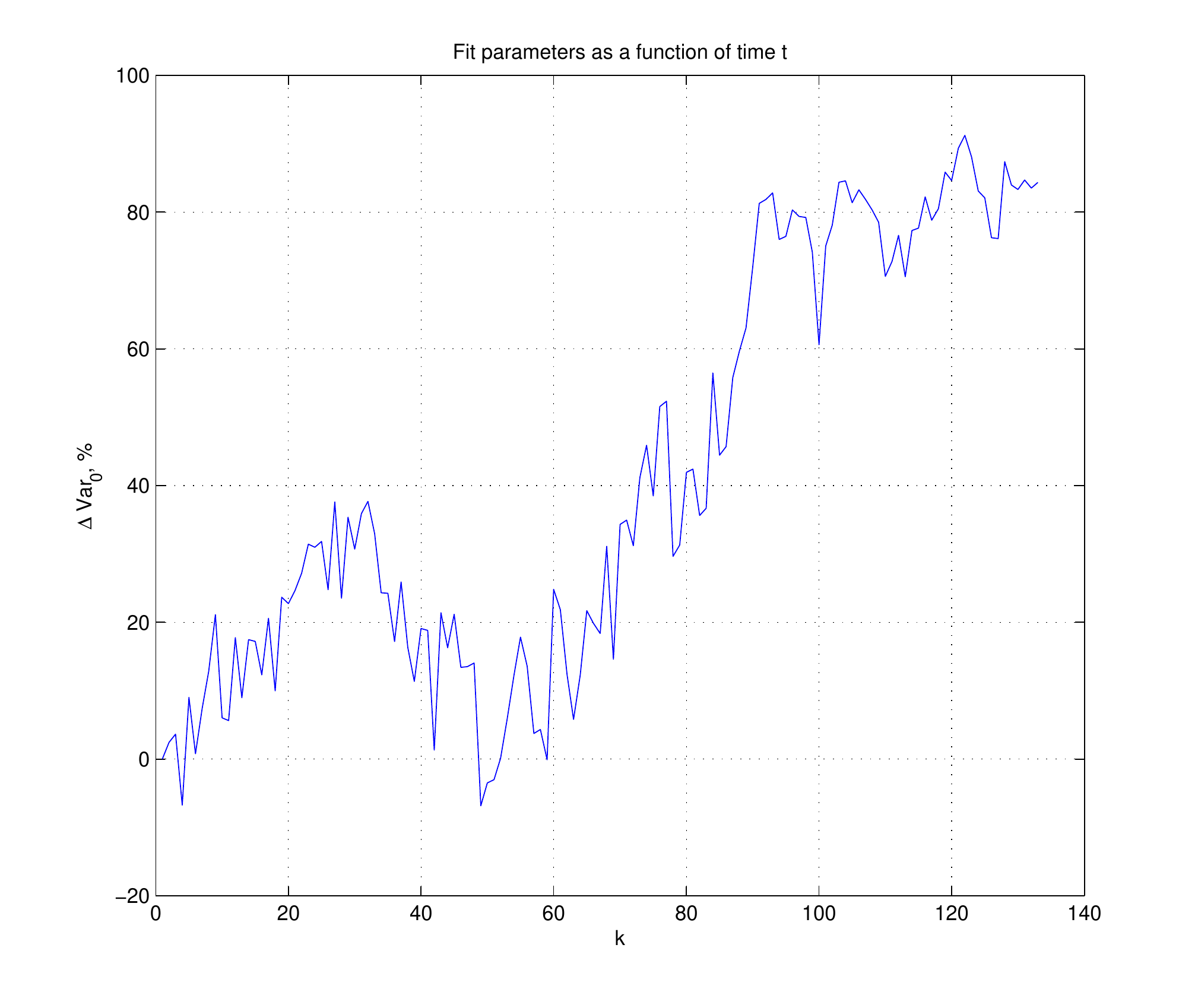}}
\caption{Sensitivity of $w_C$ to the time change.}
\label{Fig2}
\end{center}
\end{minipage}
\hspace{0.1\textwidth}
\begin{minipage}[b]{0.4\textwidth}
\begin{center}
\fbox{\includegraphics[width=3in, height = 3in]{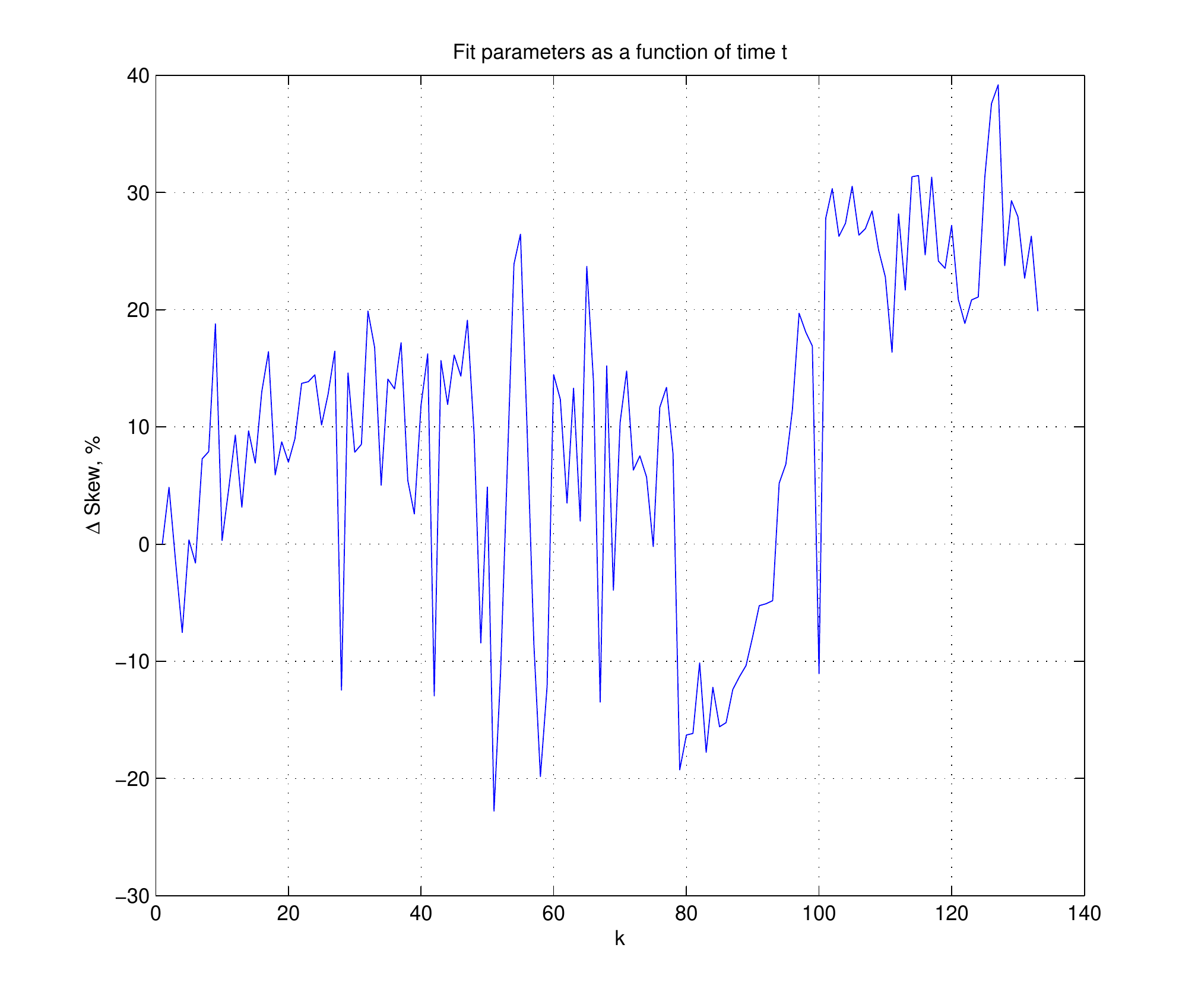}}
\caption{Sensitivity of $\cal S$ to the time change.}
\label{Fig3}
\end{center}
\end{minipage}
\end{figure}

\begin{figure}[H]
\begin{minipage}[b]{0.4\textwidth}
\begin{center}
\fbox{\includegraphics[width=3in, height = 3in]{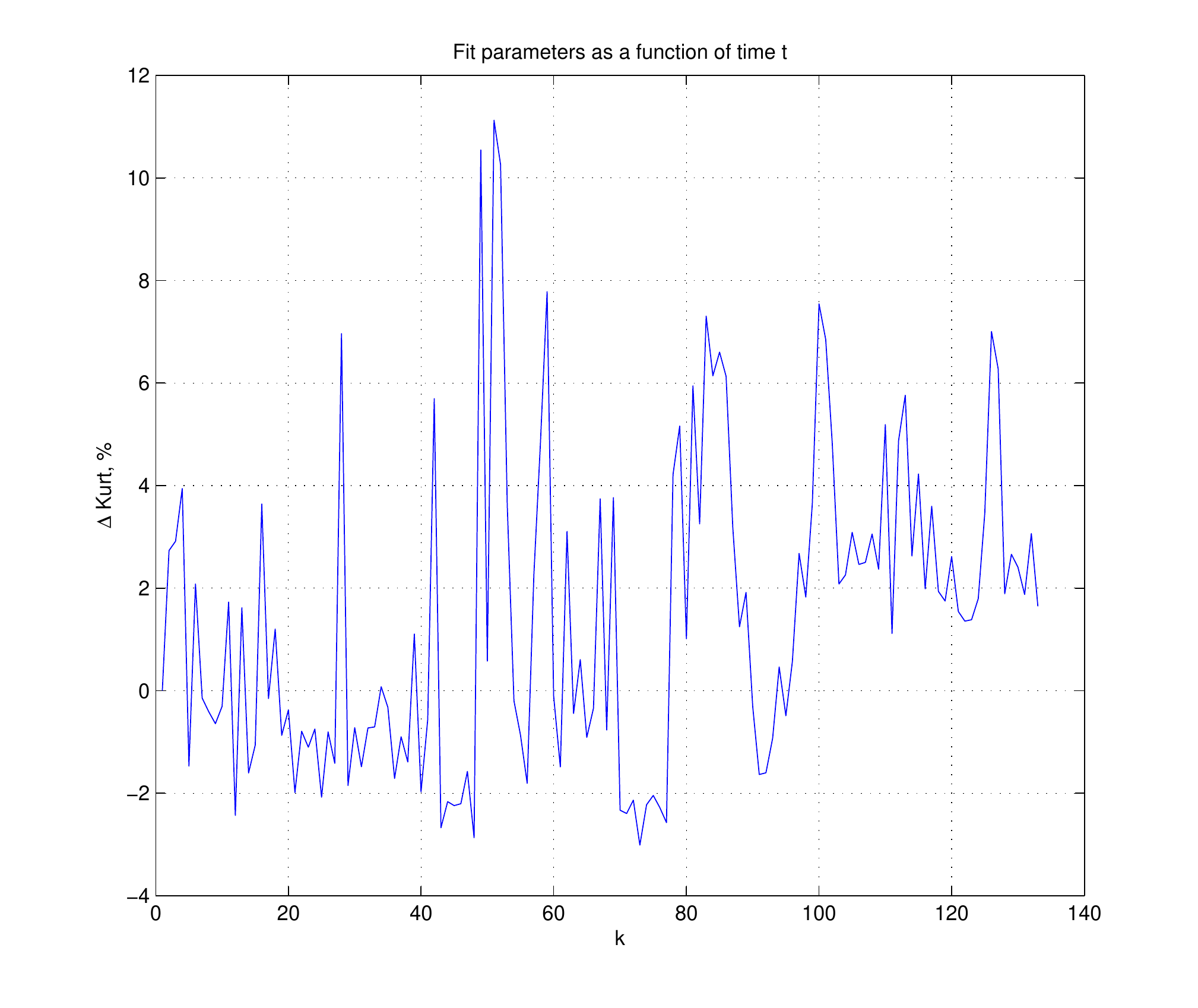}}
\caption{Sensitivity of $\cal K$ to the time change.}
\label{Fig4}
\end{center}
\end{minipage}
\hspace{0.1\textwidth}
\begin{minipage}[b]{0.4\textwidth}
\begin{center}
\fbox{\includegraphics[width=3in, height = 3in]{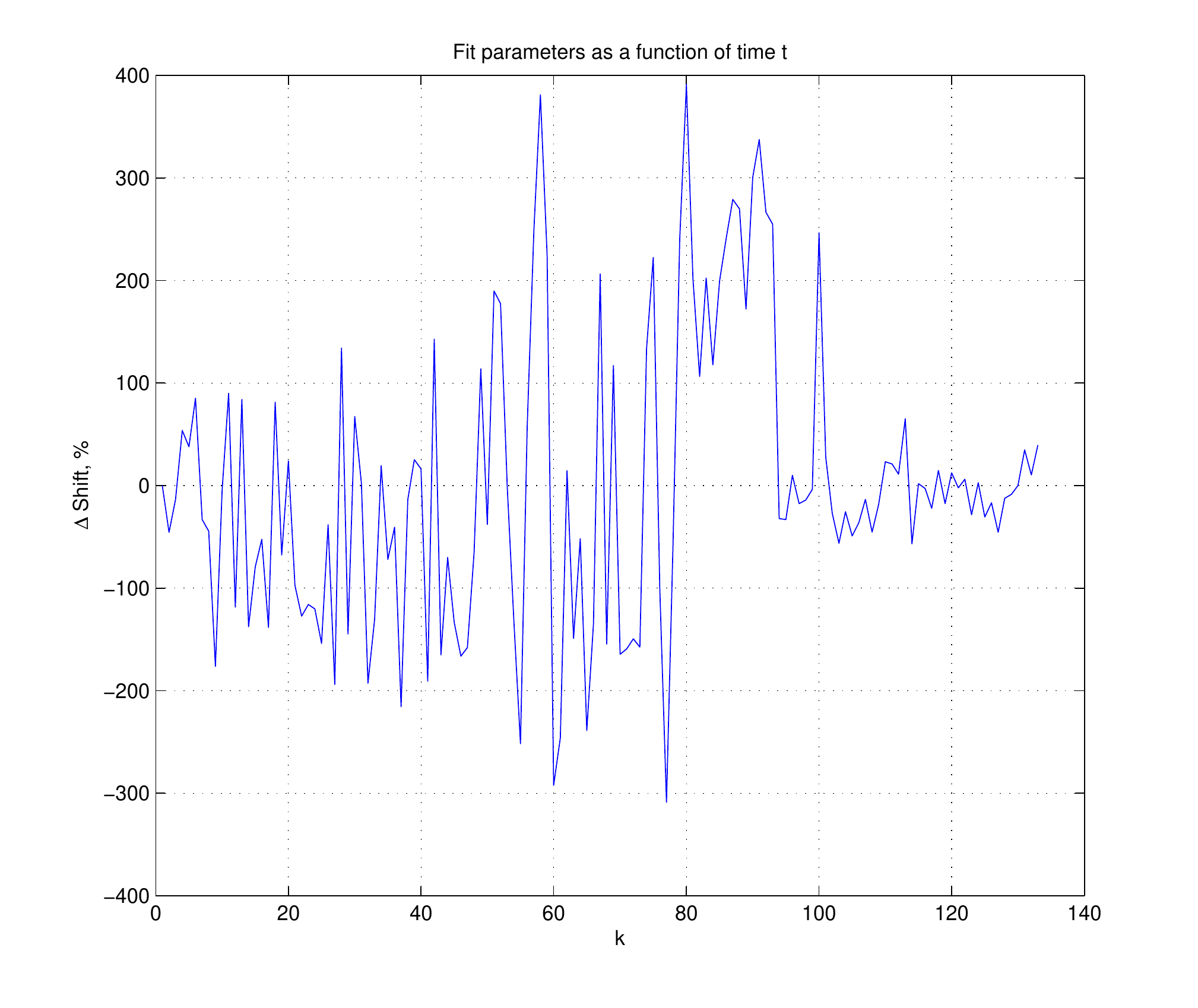}}
\caption{Sensitivity of $\cal C$ to the time change.}
\label{Fig5}
\end{center}
\end{minipage}
\end{figure}

\begin{figure}[H]
\begin{minipage}[b]{0.4\textwidth}
\begin{center}
\fbox{\includegraphics[width=3in, height = 3in]{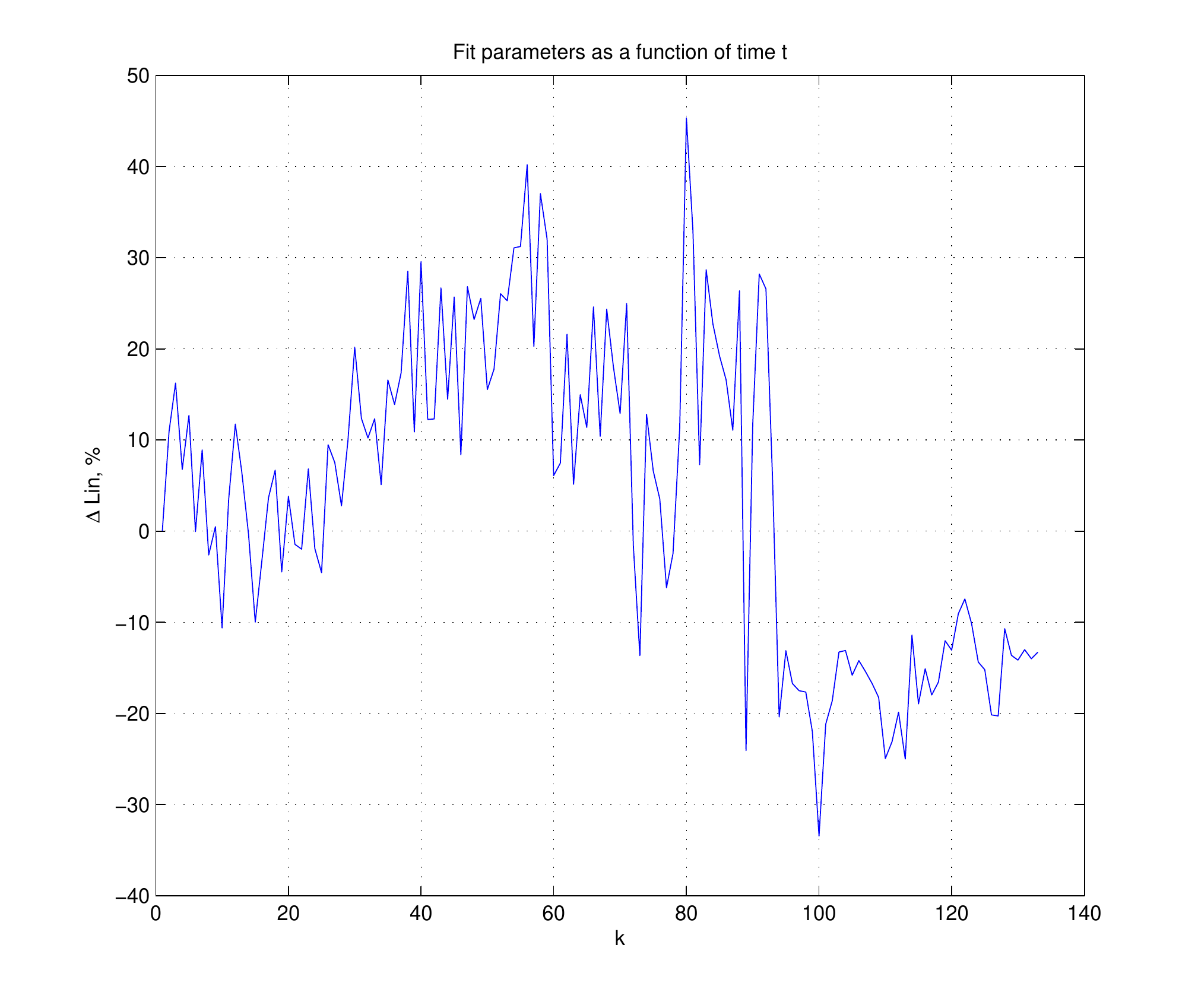}}
\caption{Sensitivity of ${\cal S}_C$ to the time change.}
\label{Fig6}
\end{center}
\end{minipage}
\hspace{0.1\textwidth}
\begin{minipage}[b]{0.4\textwidth}
\begin{center}
\fbox{\includegraphics[width=3in, height = 3in]{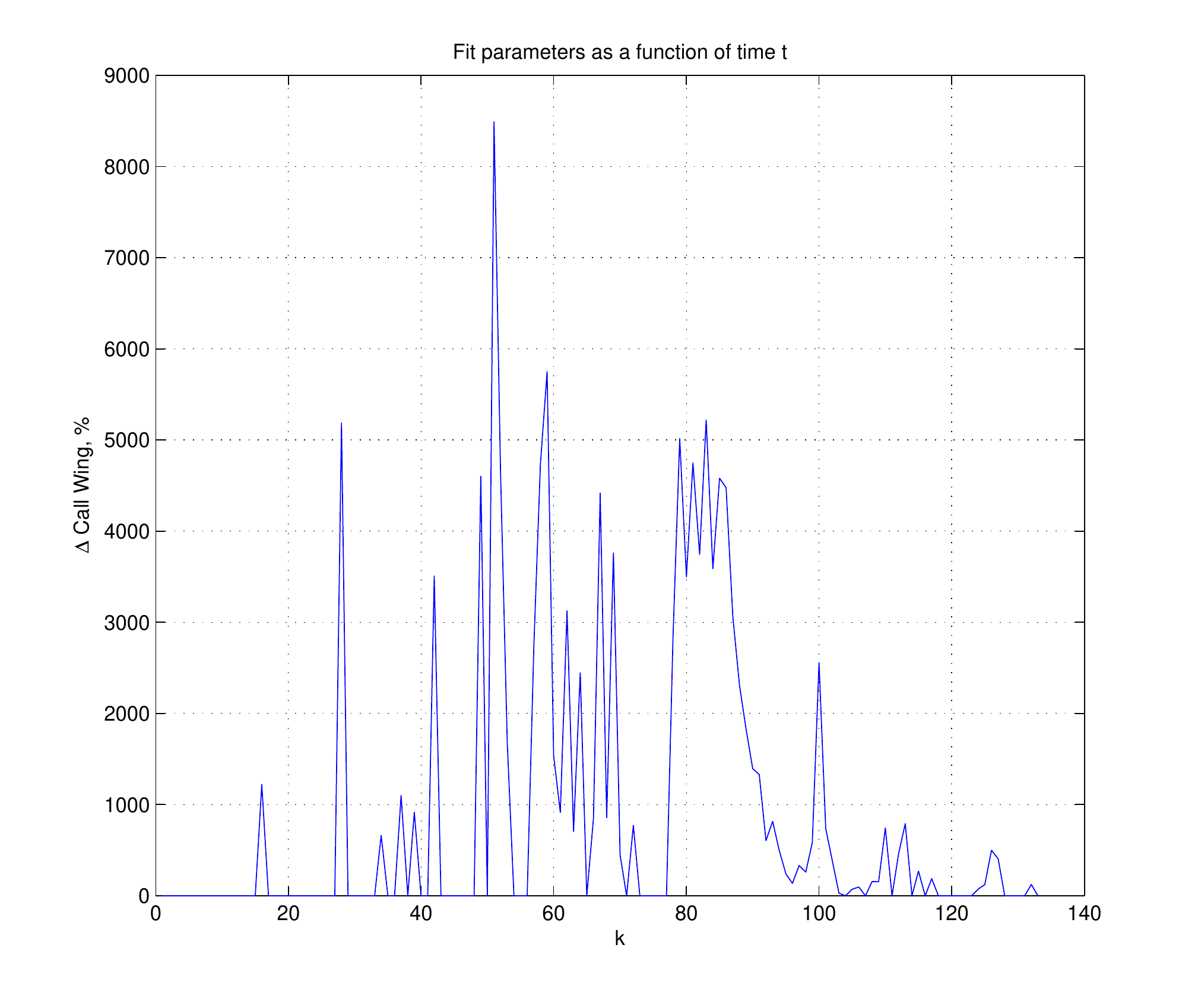}}
\caption{Sensitivity of $\beta$ to the time change.}
\label{Fig7}
\end{center}
\end{minipage}
\end{figure}

\begin{figure}[H]
\begin{center}
\fbox{\includegraphics[width=3in, height = 3in]{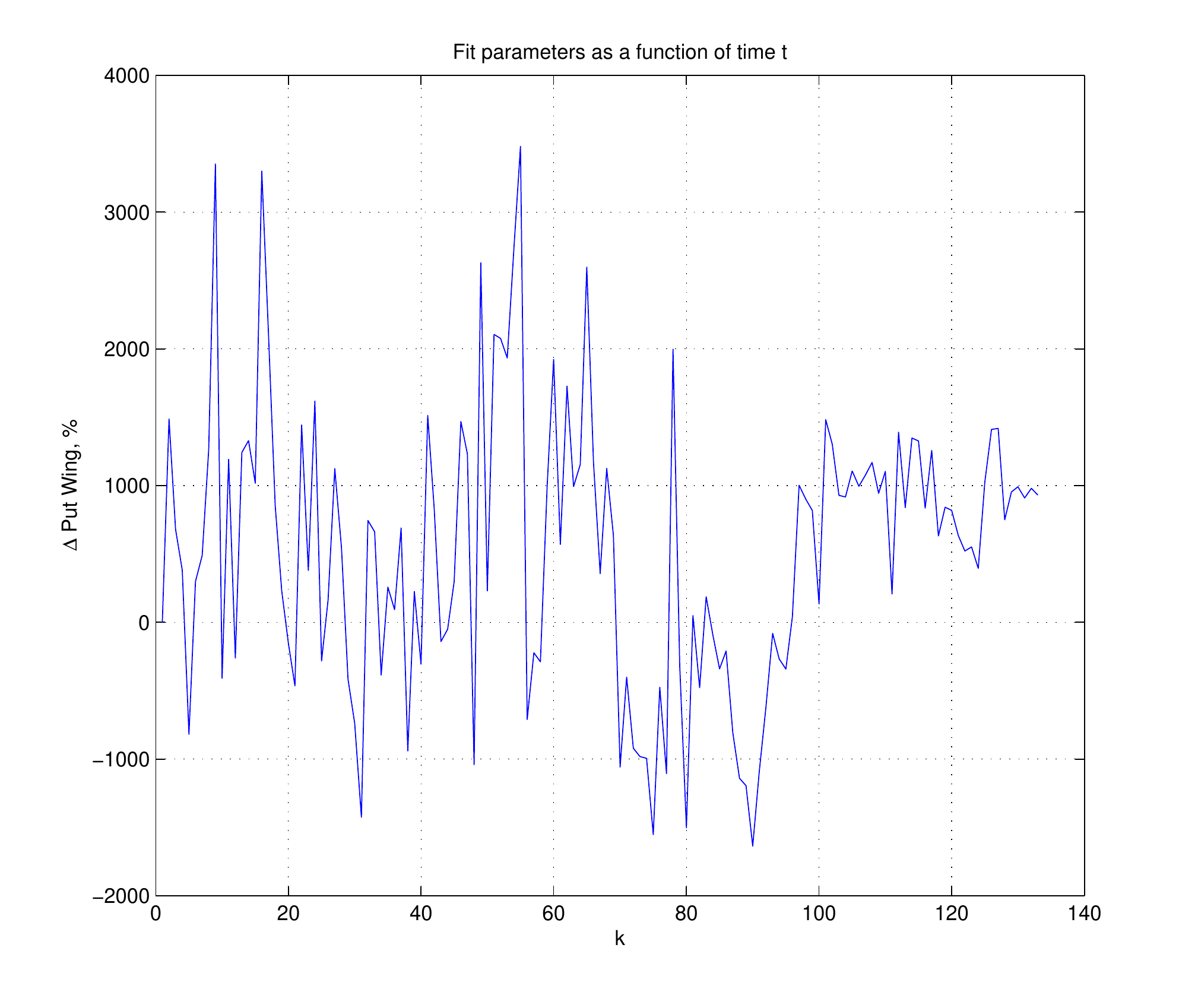}}
\caption{Sensitivity of $\alpha$ to the time change.}
\label{Fig8}
\end{center}
\end{figure}

It is interesting to see, however, how the fitting parameters behave as a function of time. In other words, what is the sensitivity of the fit to changes in time. To investigate this we use the same data on the SPX closing IVs given for 133 sequential days and plot time-series of the model parameter values. These results are given in Fig.~\ref{Fig2}-\ref{Fig8}.

As one can see, the most time sensible parameters are $\alpha$ and $\beta$. So they need to be refitted more often, probably few times a day. At the time scale of a few days other parameters change just within 10-20 \%; therefore, they could be refitted less often.

We want to emphasize that by construction our model provides just the static fit of the current market snapshot of the options IVs, and does not consider any dynamics of the IV surface. Therefore, the dependence of the model parameters on time serves just to the illustrative purposes and helps in organizing a rapid calibration procedure. This does not mean that looking at the time dependence of the model parameters one can make a predictive conclusion on how the future IV behaves with time.

\subsection{Constructing a local volatility surface} \label{example}

In this example we take data from \url{http://www.optionseducation.org} on XLF traded at NYSEArca on March 25, 2014. The spot price of the index is $S = 22.64$, the interest rate $r = 0.0148$. The option IVs are given in Tab.~\ref{TabOpt}. We take all OTM quotes and some ITM quotes which are very close to the ATM.
\begin{table}[h!]
\begin{center}
\begin{tabular}{|c|c|c|c|c|c|c|}
\specialrule{.1em}{.05em}{.05em}
T & \multicolumn{5}{c}{K, Put} & \\
\cline{2-7}
               & 18 & 19   & 20 & 21	   & 22	 & 23	\cr
\specialrule{.1em}{.05em}{.05em}

4/19/2014  &  - & 32.90 & 26.79 &  20.14 & 15.19 & 12.93 \cr
\hline
5/17/2014  & 33.27 & 26.88 & 23.08 & 18.94 & 16.12 & 13.86  \cr
\hline
6/21/2014  & 27.84 & 23.90 & 21.07 & 18.88 & 16.95 & 15.82   \cr
\hline
7/19/2014  & 26.09 & 22.81 & 20.29 & 18.13 & 16.30 & 14.93   \cr
\hline
9/20/2014  & 24.20 & 22.23 & 20.32 & 18.76 & 17.40 & 16.41   \cr
\hline
12/20/2014 & 23.75 & 22.09 & 20.67 & 19.44 & 18.36 & 17.60   \cr
\hline
\end{tabular}

\begin{tabular}{|c|c|c|c|c|c|c|c|c|}
\specialrule{.1em}{.05em}{.05em}
T & \multicolumn{7}{c}{K, Call} & \\
\cline{2-9}
               & 21 & 22   & 23 & 24 & 25 & 26 & 27 & 28 \cr
\specialrule{.1em}{.05em}{.05em}

4/19/2014  &  -    & 15.79 & 13.38 & 15.39 & - & - & - & - \cr
\hline
5/17/2014  & 16.71 & 14.48 & -     & 13.75 & - & - & - & - \cr
\hline
6/21/2014  & 16.31 & 14.78 & -     & 13.92 & 14.28 & 16.58 & - & - \cr
\hline
7/19/2014  & 16.82 & 15.24 & -     & 14.36 & 14.19 & 15.20 & - & - \cr
\hline
9/20/2014  & 17.02 & 15.84 & -     & 14.99 & 14.56 & 14.47 & 14.97 & 16.31   \cr
\hline
12/20/2014 & 17.63 & 16.61 & -     & 15.86 & 15.47 & 15.12 & 15.18 & 15.03   \cr
\specialrule{.1em}{.05em}{.05em}
\end{tabular}
\caption{XLF option IVs: C - call options, P - put options.}
\label{TabOpt}
\end{center}
\end{table}
At the overlapped strikes for calls and puts we take an average of $I_{call}$ and $I_{put}$ with weights proportional to $1 - |\Delta|_c$ and $1-|\Delta|_p$ correspondingly \footnote{By doing so we do take into account effects reported in \cite{callPutIV} that the IVs calculated from  identical  call and  put options  have  often  been  empirically  found  to  differ,  although  they  should  be  equal  in theory. However, our weights are a pure empirical rule of thumb, and more detailed investigation of this is required.}. We use the proposed parametric fit to construct the IV surface at all given expirations and strikes in a range $K \in [17,28]$ with the step $0.5$. In doing so we calibrate the first and the last term using the above described algorithm. The other terms are found on the grid by applying the arbitrage-free interpolation with $a(T) = C(K,T,I(K,T))$ and $K=17$ (in this case $I(K,T)$ is assumed to be provided in the given set of data).

\begin{figure}[!ht]
\begin{minipage}[b]{0.4\textwidth}
\begin{center}
\fbox{\includegraphics[height=2.5in, width = 3in]{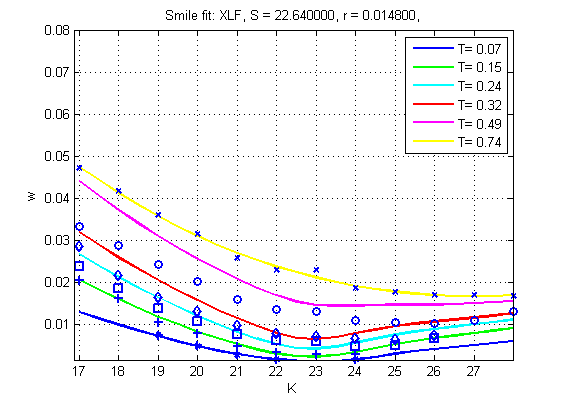}}
\caption{Term-by-term fitting of the IV surface constructed using the whole set of data in Tab.~\ref{TabOpt}.}
\label{termXLF}
\end{center}
\end{minipage}
\hspace{0.1\textwidth}
\begin{minipage}[b]{0.4\textwidth}
\begin{center}
\fbox{\includegraphics[height=2.5in, width=3in]{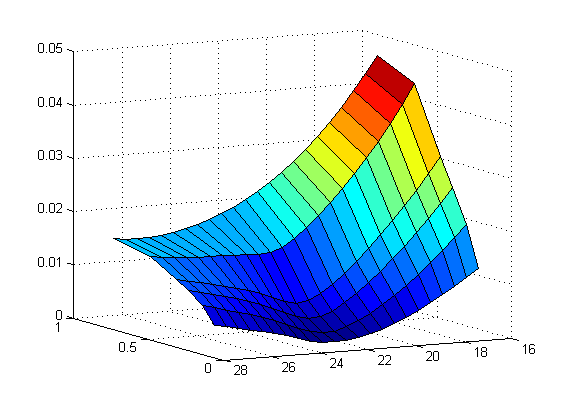}}
\caption{The IV surface obtained using the first and last terms in Tab.~\ref{TabOpt} and interpolation.}
\label{ivXLF}
\end{center}
\end{minipage}
\end{figure}

\begin{figure}[!ht]
\begin{minipage}[b]{0.4\textwidth}
\begin{center}
\fbox{\includegraphics[height=2.5in, width = 3in]{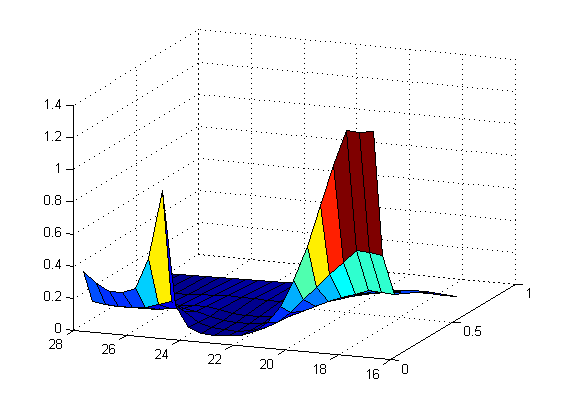}}
\caption{The local volatility surface produced from the IV surface.}
\label{lvXLF}
\end{center}
\end{minipage}
\hspace{0.1\textwidth}
\begin{minipage}[b]{0.4\textwidth}
\begin{center}
\fbox{\includegraphics[height=2.5in, width=3in]{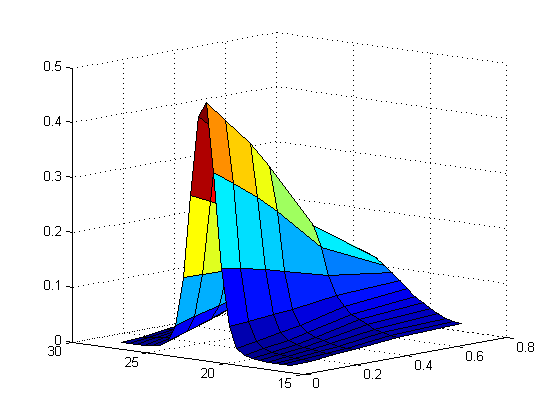}}
\caption{The density implied from the IV surface in Fig.~\ref{ivXLF}.}
\label{ivDenXLF}
\end{center}
\end{minipage}
\end{figure}

The results of this fitting are given term-by-term in Fig.~\ref{termXLF}. Accordingly, thus constructed the IV surface is represented in Fig.~\ref{ivXLF}, and the local volatility surface and the implied density obtained from the IV surface by applying the Dupire's and Breeden-Litzenberger formulas are given in Fig.~\ref{lvXLF}, \ref{ivDenXLF}.

One can see that the local volatility is positive everywhere on the grid which is provided by i) using the no-arbitrage constraints when calibrating each term, and ii) using the arbitrage-free interpolation instead of calibration for some terms.

Also the above results clearly show that calibration provides a very good fit to the market data for the first and the last term. However, for the other terms the no-arbitrage conditions could be very restrictive. Therefore, the genetic algorithm requires many evaluations of the objective function, and could be slow. In contrast, the arbitrage-free interpolation is very fast but doesn't give such a good fit to the market data.

\section{Discussion}

When fitting the IV surface we rely on raw quotes for liquid options provided by the market. Unfortunately, markets are different. For instance, in the oil market only a few strikes are traded as European listed options, usually the ATM, one ITM and one OTM option. Other strikes are traded OTC and, moreover, as Asian options. Also, the smile behavior at wings could be different for index and equity options. The variance smile could still be linear in $z$ at wings but with a very different skew. And it seems there is no a clear theoretical reason why it could not be. Therefore, if somebody has just an intuition on how the smile wings should behave, he/she could better rely on this intuition rather than on some unreliable illiquid data, and treat the latter as outliers.

Fortunately, the proposed model is able to address such an intuition by doing the following trick. Suppose that we want to have the new model for the index options smile at the call wing being as close as possible to the existing smile produced by some proven (reference) model. Then we can move the value of the call wing parameter $\beta$ into a different region by imposing a special constraint. By doing that, we make the fit a bit worse, but thus found slope (after the minimization is done) turns to be closer to the corresponding reference model skew. And we still preserve the continuity of the model.

Another issue with the model is as follows. Suppose, for a given term the number of strikes for which the market quotes are available, is less than the number of the model parameters, i.e. 7. In this case the parameterization is over determined. The no-arbitrage constraints and the asymptotic behavior of the smile help to resolve this however could not be sufficient. For instance, if only the ATM quote is liquid and available. In this situation we have either to reduce the number of parameters, or to use some tricks.

To give an example of such a trick consider the case when only a single quote $I$ corresponding to the strike $K$ is available given the time to expiration $T$. Under this situation it doesn't make sense to calibrate our parameterization to this single point. Instead, we treat the entire term as fully unknown, and find the IV values by using the arbitrage-free interpolation in time as it was described in the above. However, to fit exactly thus found IVs to the given quote we exploit the remaining flexibility in the definition of the function $a(T)$. We remind that $a(T)$ was not yet defined explicitly, and only indirectly via the condition $\dd_T a(T) > 0$. Then taking $a(T) = C(K, T, I)$ provides this inequality on one hand. On the other hand, as it could be easily checked, thus defined $a(T)$ matches exactly the given quote $C(K,T,I)$.

When 3 or 4 quotes are available for the given term, the kurtosis $\mathcal{K}$ is a natural candidate to be removed
from the parameterization. In other words, we fix $\mathcal{K} = 0$ and so reduce the total number of parameters to 6. Also, $\mathcal{S}_C$ could be the next preferable choice to omit.

As far as the relation of the proposed model to the existing ones and comparison of our results with that obtained using, e.g., the SVI model we want to underline the following. There are at least two problems that, e.g.,  quantitative analysts are dealing with pretty often, and that require knowledge of the implied volatility. One, which is important for traders and market makers, is to fit the existing market IV data, basically on a term-by-term basis, so the IV values in between of the tradable strikes could be out of their interest. This problem could be solved sufficiently well using various popular models, including SVI and the modern quadratic fit, while the proposed in this paper model also falls into this class. Then to choose an appropriate model questions about stability of the model parameters, uniqueness of the set of parameters that provide a reasonable fir to the given smile/skew should be addressed. An interesting discussion on this subject as applied to the SVI model could be found in (\cite{phFinSVI}). The SVI model often produces a non-unique set of the calibrated parameters in a sense that using various initial guesses in the calibration procedure one can get different sets of the SVI parameters that fit the given market data with almost same accuracy. This means that stability of fitting parameters could be in question. In our model to eliminate a possible instability a practical recipe is as follows. We calibrate the model for the first time, and then, when after some period of time we need to refit it, we fix one of the parameters, for instance, $\mathcal{C}$, at the previous level, thus refitting only the remaining parameters. Then in our experience the daily variations of the model parameters are pretty much suitable for traders, i.e., the fit could be treated as stable.

The second and more challenged problem is to build a local volatility function for some grid pricer given the market option quotes. This could be addressed  by first building the IV surface and then using the Dupire's formula. Here a good quality of the fit at the given quotes is not sufficient, and in addition no-arbitrage constraints in every grid point in the $(K,T)$ space as well as the correct asymptotic behavior of the smile/skew should be preserved. Under these restrictive conditions perhaps any model, including SVI, which operates just with 5 model parameters is not flexible enough to be able to meet all the constraints. Thus, it either sacrifices  by the quality of the fit, or by the no-arbitrage conditions, or by the correct asymptotic behavior in order to converge. Finally, finding the correct fit could be slow because of these limitations. As a possible resolution in this paper we demonstrate that fitting just the first and the last term and then using the arbitrage-free interpolation could be a reasonable alternative from both performance and goodness of the fit point of view. Also as our model contains more parameters, it provides an additional flexibility to better solve the above constrained optimization problem.

\begin{figure}[!ht]
\begin{minipage}[b]{0.4\textwidth}
\begin{center}
\fbox{\includegraphics[height=2.5in, width = 3in]{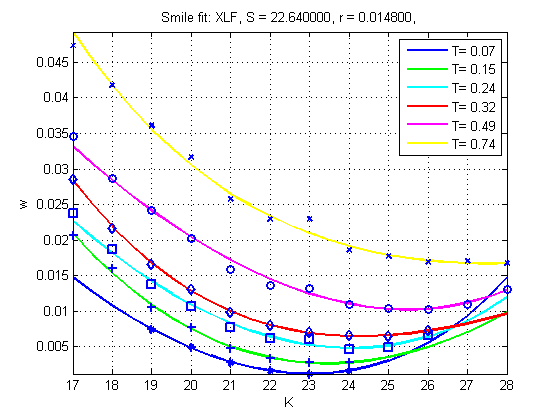}}
\caption{Term-by-term fitting of the IV surface based on the data in Tab.~\ref{TabOpt} and the SVI model.}
\label{termIV_SVI}
\end{center}
\end{minipage}
\hspace{0.1\textwidth}
\begin{minipage}[b]{0.4\textwidth}
\begin{center}
\fbox{\includegraphics[height=2.5in, width=3in]{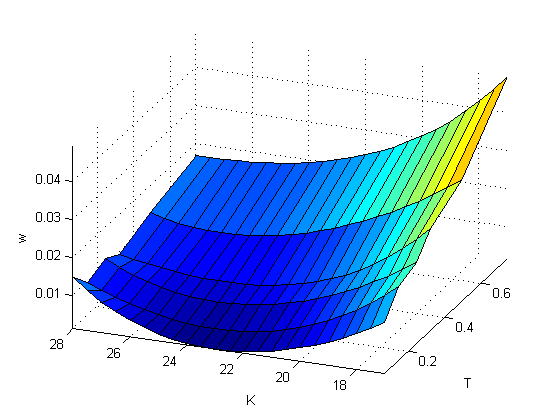}}
\caption{The local volatility surface produced from the IV surface using the SVI model.}
\label{iv_SVI}
\end{center}
\end{minipage}
\end{figure}

\begin{figure}[!ht]
\begin{minipage}[b]{0.4\textwidth}
\begin{center}
\fbox{\includegraphics[height=2.5in, width = 3in]{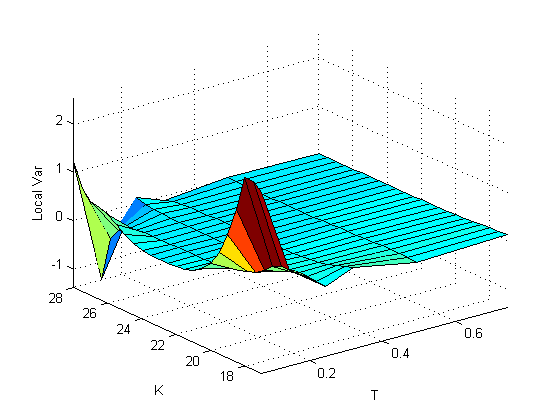}}
\caption{The local volatility surface built from the SVI IV surface.}
\label{lv_SVI}
\end{center}
\end{minipage}
\hspace{0.1\textwidth}
\begin{minipage}[b]{0.4\textwidth}
\begin{center}
\fbox{\includegraphics[height=2.5in, width=3in]{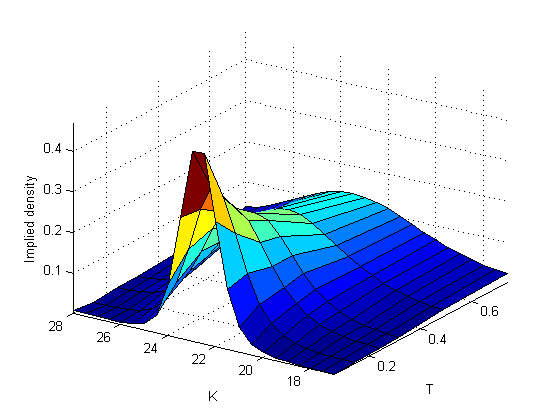}}
\caption{The density implied from the SVI IV surface shown in Fig.~\ref{iv_SVI}.}
\label{id_SVI}
\end{center}
\end{minipage}
\end{figure}

To illustrate this in Fig.~\ref{termIV_SVI}-\ref{id_SVI} we present the results of the test described in section~\ref{example}, where now instead of our model the SVI model was used. When calibrating this model in the first test the no-arbitrage constraints on a grid were not taken into  account except positivity of the total variance $w$.

The quality of the IV fit is good, however, the no-arbitrage conditions in the strike space are not validated as well as the calendar arbitrage could be observed for the OTM call options. Accordingly, the grid local volatility is negative at some strikes and expirations. However, the implied density is positive on the grid.

Surprisingly, if we take the no-arbitrage constraints into account and repeat the above test, at least in our numerical experiments the optimizer was never able to find a good fit. Moreover, thus found model parameters always produced negative local volatilities at some strikes and expirations. This justifies our hypothesis that constructing the local volatility surface on a grid by using the implied volatility surface calibrated to the market data with the SVI model could be inefficient. Most likely bad fit (and as the consequence - negative local volatilities) are observed due to insufficient flexibility of the SVI model which has only 5 parameters per one expiration term.

\section{Conclusion}

In this paper we proposed a new parametric fit of the implied volatility surface which has all the advantages that the best known models (SVI, recent versions of the quadratic fit) hold. Among those are: 1) a single regression is able to fit both skew and smile; 2) the asymptotic behavior at wings follows the existing theoretical results, i.e. the implied variance could be linear or sublinear in the normalized strike $z$; 3) we provide a calibration procedure and also interpolation/extrapolation algorithm which guarantee no arbitrage on a given grid in $(K,T)$ space; 4) the parametric function is continuous in $z$; 5) It allows a natural extension by switching from the calendar time to the business time; 6) the parametric function is fast to evaluate; 7) the local volatility surface and the implied density could be easily build from thus obtained IV surface and are arbitrage-free as well.

By its nature this model provides just the static fit of the current market snapshot of the options IVs, and does not consider any dynamics of the IV surface.

Our results demonstrate that this model could be more flexible than, e.g., SVI. We demonstrate the results of the test were our model provides good quality of the fit while preserves all necessary properties of the IV surface.
The local volatility and implied density surfaces are also constructed in this test while the SVI model experiences a problem in doing so.

\clearpage
\section*{Acknowledgments}
I thank Peter Carr, Alex Lipton, Dmitry Kreslavsky, Roza Galeeva, Antonie Kotz\'e, Lewis Biscamp and Ping Sun for fruitful discussions, and two anonymous referees whose remarks significantly improved the content of the paper.


\newcommand{\noopsort}[1]{} \newcommand{\printfirst}[2]{#1}
  \newcommand{\singleletter}[1]{#1} \newcommand{\switchargs}[2]{#2#1}

\clearpage
\appendix
\section{Finding parameters for one term} \label{ap}

To obtain the values of the smile parameters, a non-linear least square optimization is used. Every market point is taken with some weight which is usually of the following form
\begin{eqnarray}  \label{weight}
w(z) &=& \dfrac{1}{2}\left(w_c(z) + w_p(z)\right), \\
& & w_c(z) = \left(1-|\Delta_c|\right) \min \left[0.1, \left(\dfrac{z}{\sigma_{atm}}\right)^\nu\right] \nn \\
& & w_p(z) = \left(1-|\Delta_p|\right) \min \left[0.1, \left(\dfrac{z}{\sigma_{atm}}\right)^\nu\right] \nn
\end{eqnarray}

Here $\Delta_c$ and $\Delta_p$ are the market call and put deltas of the option, $\sigma_{atm}$ is the ATM market implied volatility, $\nu$ is some parameter which is typically taken to be -2 or -3. Having these weights, the following optimization problem was solved to obtain parameters of the fit
\begin{equation}  \label{opt}
\min_{p_1...p_7} \sum_{i=1}^N W_i(z) \left[ w_m(z_i) - w(z_i, p_1...p_7) \right]^2,
\end{equation}

\noindent  where $N$ is the total number of the raw option data, $W_i(z)$ is the weight of the $i$th point, $w_m$ is the market total implied variance of the data, $\nu_i, i=1,7$ are the parameters of the model.

This minimization problem is solved under the whole bunch of no-arbitrage constraints discussed in the previous section. The no-arbitrage constraints are checked at every node on the grid $\mathbb{G}$, while the asymptotic slope is checked at two edge points on the grid for every time slice.

Further on, we calibrate all terms, provided as an input, that contain at least a single data point, by using bootstrap, i.e. term by term. We start with reordering all the market data in the ascending order, and then proceed with fitting the shortest term at $T = T_1$. Then the next term at $T = T_2$ is fitted, etc. To solve this optimization problem  we use a genetic algorithm implemented in \cite{CMAES} which we updated with allowance for the equality and inequality constraints. This algorithm guarantees finding a global minimum. A typical time necessary to get the values of the parameters for one term in C++ is about 0.5 secs with the maximum number of function evaluations set to $10^4$. Based on our experiments this value provides a very good fit, while it could be  lowered to get a better performance. Note that as our algorithm belongs to the class of evolutionary optimization (\cite{Simon2013}) it is very suitable for parallelization.

We also need to underline that the above minimization problem is solved at points $z_i, \ i=1...s$ where for the given term $T_j$ the market prices are available at $s$ strikes $\hat{K}_i, \i=1...s$ such that $z_i = (\log \hat{K}_i/F(T_j))/\sqrt{T_j}, \ i=1...s$. However, the no-arbitrage constraints are checked at another set of points: that ones that belong to the $\mathbb{G}$ grid. According to the definition of $\mathbb{G}$ these points are $z_i = (\log(K_i)/F(T_j))/\sqrt{T_j}, \ i=1...M$. By construction the implied volatilities obtained on the grid nodes are arbitrage-free.

\paragraph{Smart initial guess.}
When calibrating every term at the beginning we use a special algorithm to provide a good initial guess. The idea behind this algorithm is that as follows from \eqref{fit} $w_{zz}(z) = 0$ at the point $z=\mathcal{C}$. Therefore, one can look at the input IVs in the $z$ space, compute the second derivative and find where it vanishes. In case the second derivative is positive everywhere, as the value of $\mathcal{C}$ one can take such $z$ where the IV is minimal among all values belonging to this term. This construction also works well when the IV surface has a skew, not a smile, which is typical for index options.

Given $\mathcal{C}$ other parameters could be obtained relatively straightforward. Indeed, from \eqref{fit}
\begin{equation} \label{sig}
w_C = w(z)|_{z=C}, \qquad {\cal S}_C = w_z(z)|_{z=C}.
\end{equation}
Now denote
\[ \kappa(y) = \frac{1}{y \tanh(p y) \sqrt{T}} \left(w(y+\mathcal{C}) - w(\mathcal{C}) - \mathcal{S}_C \frac{y}{1+y^2}\right), \qquad y \equiv z-\mathcal{C}, \]
so based on \eqref{sig} $\kappa(y)$ is a known function of $y$. Accordingly \eqref{fit} could be re-written in the form
\begin{equation} \label{lastPars}
{\cal S} Y(y) + {\cal K} Y^2(y) = \kappa(y).
\end{equation}
To find the initial guess for the remaining parameters $\alpha, \beta, {\cal S}, {\cal K}$ we need 4 additional market IVs. At least two of them should lie on the different sides of the IV curve with regard to the point $y=0$. As an example, consider three points $y_1 > y_2 > y_3 > 0$. Then, $Y(y)$ in \eqref{lastPars} is defined via parameter $\beta$, see \eqref{fit}. Using \eqref{lastPars} with $y_1$ and $y_2$ we find
\begin{equation} \label{SK}
\mathcal{S} = \frac{\kappa(y_1) Y^2(y_2) - \kappa(y_2) Y^2(y_1)}
{Y(y_1) Y^2(y_2) - Y(y_2) Y^2(y_1)}, \qquad
\mathcal{K} = \frac{\kappa(y_2) Y(y_1) - \kappa(y_1) Y(y_2)}
{Y(y_1) Y^2(y_2) - Y(y_2) Y^2(y_1)}.
\end{equation}
Now using the point $y_3$ we numerically solve the equation \eqref{lastPars} with regard to $\beta$.

Since parameters $\mathcal{S}, \mathcal{K}$ are already found, the last point $y_4 < 0$ could be used together with \eqref{lastPars} to numerically solve for $\alpha$. This finalizes computation of the initial guess.

In case the input data points are located as $y_1 > y_2 > 0 > y_3 > y_4$ the easiest way is to add an extra point to the negative $y$ by using interpolation, and then remove one point, e.g., $y_2$ from the positive $y$, thus getting back to the previous case.

\end{document}